\documentclass[letterpaper,11pt]{article}

\usepackage{jcappub} 

\usepackage{tcolorbox}
\usepackage{latexsym}
\usepackage{amsfonts}
\usepackage{mathrsfs}
\usepackage{amssymb,amscd}
\usepackage[all]{xy}
\usepackage{amsmath}
\usepackage{amsthm}
\usepackage{fancyhdr}
\usepackage{fancybox}
\usepackage{xcolor}
\usepackage{hyperref}
\usepackage{verbatim}
\usepackage{makeidx}
\usepackage{indentfirst}
\usepackage[english]{babel}
\usepackage{graphicx}
\usepackage{listings}
\usepackage{subfigure}
\usepackage{multirow}
\usepackage{lscape}
\usepackage{physics}
\usepackage{amsmath}
\usepackage{framed}
\usepackage{bbold}
\usepackage{siunitx}
\usepackage{comment}
\usepackage{cases}
\usepackage{booktabs}

\usepackage{bm}
\numberwithin{equation}{section}

\usepackage{xparse}

\usepackage{hyperref}
\hypersetup{
    colorlinks=true,
    linkcolor=blue,
    filecolor=magenta,      
    urlcolor=cyan,
    pdftitle={Overleaf Example},
    pdfpagemode=FullScreen,
    }

\newcommand{\kvec}{\bm{k}}

\def\c{{\boldsymbol c}}

\def\d{{\rm d}}

\def\x{{\boldsymbol x}}
\def\y{{\boldsymbol y}}

\def\btheta{{\boldsymbol{\theta}}}

\def\bepsilon{\boldsymbol\epsilon}

\title{Field-level Reconstruction from Foreground-Contaminated 21-cm Maps}
\author{Shu-Fan Chen$^{a}$, Kai-Feng Chen$^{b,c}$, and Cora Dvorkin$^{a}$} 
\affiliation{$^{a}$Department of Physics, Harvard University, Cambridge, MA 02138, USA\\
  $^b$Department of Physics, Massachusetts Institute of Technology, Cambridge, MA 02139, USA\\
  $^c$MIT Kavli Institute, Massachusetts Institute of Technology, Cambridge, MA 02139, USA}
\emailAdd{shufan\_chen@g.harvard.edu
 }
\emailAdd{kfchen@mit.edu}

\abstract{Current and upcoming 21-cm experiments will soon be able to map 21-cm spatial fluctuations in three dimensions for a wide range of redshifts. However, bright foreground contamination and the nature of radio interferometry create significant challenges, making it difficult to access rich cosmological information from the Fourier modes that lie within the ``foreground wedge''. In this work, we introduce two approaches aiming to reconstruct the full 21-cm density field, including the missing modes in the wedge: (a) a field-level inference under an effective field theory (EFT) framework; (b) a diffusion-based deep generative model trained on simulations. Under the EFT framework, we implement a fully differentiable forward model that maps the initial conditions of matter fluctuations to the observed, foreground-filtered 21-cm maps. This enables a gradient-based sampler to simultaneously sample the initial conditions and bias parameters, allowing a physically motivated mode reconstruction. Alternatively, we apply a variational diffusion model to perform 21-cm density reconstruction at the map level. Our model is trained on semi-numerical simulations over a wide range of astrophysical parameters. Our results from both approaches should provide improved cosmological constraints from the field level and also enable cross-correlation between experiments that have little or no overlapping modes.}

\begin{document}
\maketitle
\flushbottom

\section{Introduction}
The 21-cm line from neutral hydrogen is a promising probe for tracing the large-scale structure (LSS) across a large volume of the unexplored universe \cite{Furlanetto2006:Review, Pritchard2012:Review}. This isolated and highly forbidden transition in the radio band provides a unique opportunity to map the universe in three dimensions, as the observed frequency of the 21-cm line can be directly translated into the redshift (distance) of its source. Many radio interferometers have been built or proposed to detect the spatial fluctuation of the 21-cm signal. These include experiments at lower redshift (higher frequency) targeting the baryon acoustic oscillations (BAO) in the post-reionization universe \cite{CHIME2022:Overview, MeerKAT2016:MeerKLASS, CHORD2019:Overview, HIRAX2022:Overview, SKA2015:HI_IM} and at higher redshift (lower frequency) aiming for the Epoch of Reionization or the Cosmic Dawn \cite{PAPER2010:Overview, GMRT2013:21cmLimit, GMRT2017:uGMRT_Overview, MWA2013:PhaseI_Overview, MWA2018:PhaseII_Overview, LOFAR2013:Overview, NenuFAR2012:Overview, HERA2017:PhaseI_Overview, Berkhout2024:HERA_PhaseII, LWA2019:21cmLimit, SKA2015:EoR}. Successful observations of such a signal will provide rich astrophysical and cosmological information, allowing us to better understand the dark sector of our universe and unveil the environment for star and galaxy formation \citep{Chang2008:BAO_IM, Morales2010:Review, Bull2015:21cmIM, Mesinger2016:Review}. 

The most significant barrier toward a successful measurement of the 21-cm signal lies in the substantial foreground contamination, as the radio emission from both galactic and extragalactic sources can be many orders of magnitude brighter than the cosmological signal \cite{de_Oliveira-Costa2008:GSM, Zheng2017:GSM, Hurley-Walker2017:GLEAMI, Hurley-Walker2019:GLEAMII}. Although astrophysical foreground emission is expected to be spectrally smooth, the chromatic instrument response of a radio interferometer usually causes the foreground to contaminate a large area in Fourier space (known as the foreground wedge \cite{Datta2010:FGwedge, Parsons2012:delay_spectrum_wedge, Vedantham2012:image_wedge, Trott2012:wedge, Morales2012:wedge, Hazelton2013:wedge, Thyagarajan2013:wedge, Liu2014:EoR_WindowI}). While many methods have been developed to explicitly model and subtract the foregrounds \cite{Datta2010:FGwedge, Liu2011:QE, Chapman2012:FastICA, Ghosh2015:Bayesian_21cm, Ewall-Wice2021:DAYNENU, Mertens2018:GPR, Kern2021:GPR, Chege2022:MWA_foreground, Wang2024:Foreground_subtraction, Mertens2024:GPR}, accessing large-scale 21-cm fluctuations remains challenging. Moreover, foreground subtraction becomes even more daunting in the presence of uncertainties in instrument response and systematic effects \cite{Barry2016:CalibrationError, Patil2016:CalibrationError, Ewall-Wice2017:CalibrationError, Bryne2019:CalibrationError, Mouri_Sardarabadi2019:CalibrationError, Neben2016:BeamErrorMWA, Joseph2018:BeamError, Ansah-Narh2018:BeamPerturbation, Orosz2019:BeamVariation, Joseph2020:BeamVariation, Choudhuri2021:BeamVariation, Kim2022:BeamPerturbation, Ewall-Wice2016:Limit_Relection, Kern2019:Relection_Model, Ung2020:MWA_Coupling_Sim, Josaitis2022:MutualCoupling, Rath_Pascua2024:Mutual_Coupling, Chen2025:RFI}. The difficulties in accessing these large-scale modes not only reduce the cosmological sensitivity but also significantly limit the potential of cross-correlating the 21-cm signal with other tracers.

Rather than modeling and directly removing the foregrounds, some approaches adopt a more aggressive strategy by setting all modes within the foreground wedge to zero and reconstructing these modes using only the cleanly observed ones. This reconstruction leverages correlations between long- and short-wavelength modes of tracers. Quadratic or fossil estimators have been developed to exploit this correlation, using a fiducial bispectrum derived from perturbation theory~\cite{Darwish:2020prn}. However, as demonstrated in Ref.~\cite{Wang:2023lvt}, accurate mode reconstruction requires a precise bispectrum model for the quadratic estimator kernel to avoid introducing biases. To move beyond a tree-level bispectrum model, Ref.~\cite{Modi:2019hnu} investigates a field-level approach with a Lagrangian bias model extending to quadratic order for the low-redshift ($z=2-6$) post-reionization 21-cm field. Using a quasi-Newtonian optimizer, they directly obtain the maximum a posteriori (MAP) estimate based on given bias parameters. 

Field-level inference has been successfully applied to a wide range of tracers~\cite{Nguyen:2024yth,Jasche:2012kq,Kitaura:2012tu,Wang:2014hia,Jasche:2014vpa,Ata:2014ssa,Wang:2016qbz,BOSS:2016ipy,Modi:2018cfi,Jasche:2018oym,Lavaux:2019fjr,Schmidt:2018bkr,Cabass:2019lqx,Cabass:2020nwf,Cabass:2020jqo,Elsner:2019rql,Schmidt:2020tao,Modi:2022pzm,Dai:2022dso,Charnock:2019rbk,Doeser:2023yzv,Kostic:2022vok,Bayer:2023rmj,Leclercq:2021ctr,Andrews:2022nvv,Porqueres:2023drp,Beyond-2pt:2024mqz,Sullivan:2024jxe,Taruya:2021jhg}, but it has yet to be applied to tracers from the Epoch of Reionization. This opportunity emerged from recent theoretical developments~\cite{Qin:2022xho,McQuinn:2018zwa}, which demonstrated that the 21-cm field during the Epoch of Reionization—observable through the Hydrogen Epoch of Reionization Array (HERA) and future experiments—can be effectively modeled within the framework of effective field theory (EFT). Moreover, Ref.~\cite{Qin:2025olv} showed that small-scale 21-cm fluctuations encode significant information about large-scale structures, and that the EFT formalism enables scaling hydrodynamical simulations to large volumes with $\mathcal{O}(10\%)$ accuracy. These insights motivate the use of EFT in combination with quadratic estimators~\cite{Darwish:2020prn} to reconstruct long-wavelength density modes that are obscured by the foreground wedge, using only the observable modes outside of it~\cite{Qin:2025lyy}. In this work, we extend these developments in several key directions. Most importantly, we do not assume prior knowledge of the matter density at the field-level. Instead, we jointly infer both the initial conditions (ICs) and the bias parameters from the foreground-filtered 21-cm field by directly exploring the full posterior. To make this high-dimensional inference tractable, we employ a gradient-based sampler that efficiently navigates the joint parameter space. 

In addition to field-level approaches using perturbation theory, deep learning methods have also been explored for density reconstruction, allowing us to reach smaller scales where perturbation theory alone breaks down, provided small-scale physics is accurately simulated. Deep learning can learn complex, high-dimensional posterior distributions without explicitly modeling (as required in perturbative methods). Variational diffusion models~\cite{2021arXiv210700630K} have been applied to mock galaxy surveys to infer dark matter density fields in two or three dimensions within the CAMELS simulation suite~\cite{Ono:2024jhn,Park:2023ync,park2024d,Legin:2023jxc}. Convolutional Neural Networks (CNNs) and U-Nets have also been applied in several works for foreground removal~\cite{Kennedy:2023zos,Li:2019znt,Makinen:2020gvh,Gagnon-Hartman:2021erd,Bianco:2023eec,Bianco:2024jhe}. Additionally, a stochastic interpolant~\cite{Sabti:2024jff}---a hybrid of diffusion models and normalizing flows---has shown promise for reconstructing wedge-filtered regions in three-dimensional high-redshift 21-cm lightcones centered at $z \approx 10$, trained on data from semi-analytical simulations. Building on these approaches, we extend this work in two key directions. First, we apply the variational diffusion model to reconstruct the wedge-filtered 21-cm field from the Epoch of Reionization, incorporating observational noise. Second, we analyze how the correlation coefficient between the true and reconstructed fields compares with results from EFT-based field-level inference, demonstrating that both methods offer comparable wedge recovery capabilities.

This paper is organized as follows: In Section~\ref{sec:forwardModel}, we describe the modeling of the 21-cm field within the EFT framework. In Section~\ref{sec:fieldLevelInferenceWithMCLMC}, we detail our field-level inference using a gradient-based sampling algorithm and present a consistency test. In Section~\ref{sec:diffusionBasedGenerativeModel}, we turn to the deep diffusion model and outline our findings. Finally, in Section~\ref{sec:comparisonDiscussions}, we compare and discuss the results of both methods. Throughout this work, we adopt the best-fit cosmological parameters from Planck 2018~\cite{Planck:2018vyg} as our fiducial model. Specifically, we set the matter density to $\Omega_m = 0.31$, the baryon density to $\Omega_b = 0.049$, the spectral index to $n_s = 0.96$, the Hubble constant to $h = 0.67$, and the amplitude of matter fluctuations to $\sigma_8 = 0.81$.

\section{Modeling and Input Data}\label{sec:forwardModel}

\subsection{21-cm Cosmology \& Mock Observations}
The main observable in 21-cm cosmology is the absorption or emission of the 21-cm line relative to the background radio emission (predominantly from the cosmic microwave background). In the Rayleigh-Jeans limit, this brightness contrast can be quantified by the brightness temperature $\delta T_\mathrm{b}$, which can be expressed as \cite{Furlanetto2006:Review}
\begin{align}
    \delta T_b \approx 27 \,{\rm mK} (1+\delta_b)x_{\rm HI}\left(1-\frac{T_{\rm CMB}(\nu)}{T_{\rm spin}}\right)\left(\frac{\Omega_b h^2}{0.023}\right)\sqrt{\left(\frac{1+z}{10}\right)\left(\frac{0.15}{\Omega_m h^2}\right)}\left(\frac{H(z)/(1+z)}{\d v_\|/\d r_\|}\right)\,,
\end{align}
where $\delta_b$ is the baryon overdensity field, $x_{\rm HI}$ is the fraction of hydrogen that is neutral, $T_{\rm CMB}(\nu)$ is the {\color{black} temperature of the cosmic microwave background (CMB)} at frequency $\nu$, $\d v_\|/\d r_\|$ is the gradient of the proper velocity along the line-of-sight direction, and $T_{\rm spin}$ is defined as the ratio of the occupancy of the spin-1 and spin-0 ground states of the neutral hydrogen:
\begin{align}
    \frac{n_1}{n_0} = 3\exp\left(-T_* / T_{\rm spin}\right)\text{  with $T_*=0.0681$ K}\,.
\end{align}

In this work, we use \texttt{21cmFAST}\footnote{\href{https://21cmfast.readthedocs.io/en/latest/}{\color{black} https://21cmfast.readthedocs.io/en/latest/}}\cite{Mesinger2011:21cmFAST} to simulate the brightness temperature field. \texttt{21cmFAST} utilizes the second-order Lagrangian perturbation theory (2LPT) to generate the underlying density field, and uses an excursion-set formalism \cite{Furlanetto:2004sim} to simulate the process of reionization. The main ansatz of \texttt{21cmFAST} is that the heating and reionization of the intergalactic medium (IGM) are dominated by the astrophysics of the first galaxies. We adopt the eight-parameter model proposed in Ref. \cite{Park:2018ljd} to describe the properties of these galaxies. Furthermore, {\color{black} we choose to focus on the Epoch of Reionization ($z \lesssim 10)$---a time when the universe is sufficiently heated. Hence, we assume $T_{\rm spin} \gg T_{\rm CMB}$ at these redshifts and ignore the galaxy X-ray properties.} The resulting {\color{black} six} parameters and their values adopted in this work are summarized in \autoref{tab:21cmFASTTrainingParamRange}.

\begin{table}[htbp]
    \centering
    \small
    \renewcommand{\arraystretch}{1.2} 
    \begin{tabular}{l p{6.5cm} c c} 
        \hline
        Parameter & Description & Flat Prior & Fiducial Value \\
        \hline
        $\log_{10}(f_{*,10})$ & Fraction of gas in stars for $10^{10}\,M_\odot$ haloes & $[-1.34,\,-1.06]$ & $-1.30$\\
        $\alpha_*$ & Power-law index of fraction of galactic gas in stars as a function of halo masses & $[0.41,\,0.53]$ & $0.5$\\
        $\log_{10}(f_{\rm esc,10})$ & Fraction of ionizing photons escaping into the IGM for $10^{10}\,M_\odot$ haloes & $[-1.28,\,-0.94]$ & $-1$\\
        $\alpha_{\rm esc}$ & Power-law index of the escape fraction as a function of halo mass & $[-0.66,\,-0.34]$ & $-0.3$\\
        $\log_{10}(M_{\rm turn})\,[M_\odot]$ & Turn-over mass for quenching of star formation in haloes & $[8.53,\,8.95]$ & $8.70$\\
        $t_*$ & Fractional characteristic time-scale defining the star formation rate of galaxies & $[0.40,\,0.77]$ & $0.5$\\
        \hline
    \end{tabular}
    \caption{ {\color{black}Tunable parameters in {\tt 21cmFAST} and the corresponding fiducial values adopted in this work. Ranges of prior values are given here to generate training dataset for the diffusion model discussed in Sec.\,\ref{sec:diffusionBasedGenerativeModel}. The flat priors are based on the 68\% credible range estimated in Ref.~\cite{Park:2018ljd}, which combines constraints from \textit{Hubble Space Telescope} luminosity functions and the anticipated 21-cm signal from HERA.} }
    \label{tab:21cmFASTTrainingParamRange}
\end{table}

To simulate the impact of bright foreground emission at each redshift, the brightness temperature field is further filtered to mimic the process of removing foreground contamination in real data. First, we transform the three-dimensional brightness temperature field to Fourier space to obtain $\delta T_\mathrm{b}(\kvec)$, with $\kvec$ being the three-dimensional wave vector. It is useful to further decompose the wave vector into $\kvec \equiv (\kvec_\perp, k_\parallel)$ with $\kvec_\perp$ representing the Fourier dual of the spatial coordinates perpendicular to the line of sight and $k_\parallel$ denoting the Fourier modes parallel to the line of sight. If foreground emission is spectrally smooth, only a small set of Fourier modes at low $k_\parallel$ would be contaminated. However, the chromatic response of a radio interferometer introduces mode-mixing between $\kvec_\perp$, {\color{black}$k_\parallel$}, and the foreground contamination spans over a large area in Fourier space known as the $\textit{foreground wedge}$. In this work, for each $\delta T_\mathrm{b}(\kvec)$ generated from a simulation, we consider the $\textit{observed}$ brightness temperature field $\delta T_\mathrm{b}^\mathrm{obs}(\kvec)$ to be \cite{Liu2014:EoR_WindowI}
\begin{equation}
\label{eq:foreground_filter}
    \delta T_\mathrm{b}^\mathrm{obs}(\kvec) \equiv \begin{cases}
        & 0 \quad\textrm{if }|k_\parallel| \leq \frac{D_\mathrm{c}(z)H(z)}{c(1+z)}|\kvec_\perp|\\
        & \\
        & \delta T_\mathrm{b} (\kvec) + n(\kvec)\quad\textrm{otherwise}
    \end{cases},
\end{equation}
where $D_\mathrm{c}(z)$ is the comoving line-of-sight distance, $H(z)$ is the Hubble parameter as a function of redshift, $c$ is the speed of light, and $n$ describes the instrumental noise. 

\begin{figure}[htb]
    \centering
    \includegraphics[width=\linewidth]{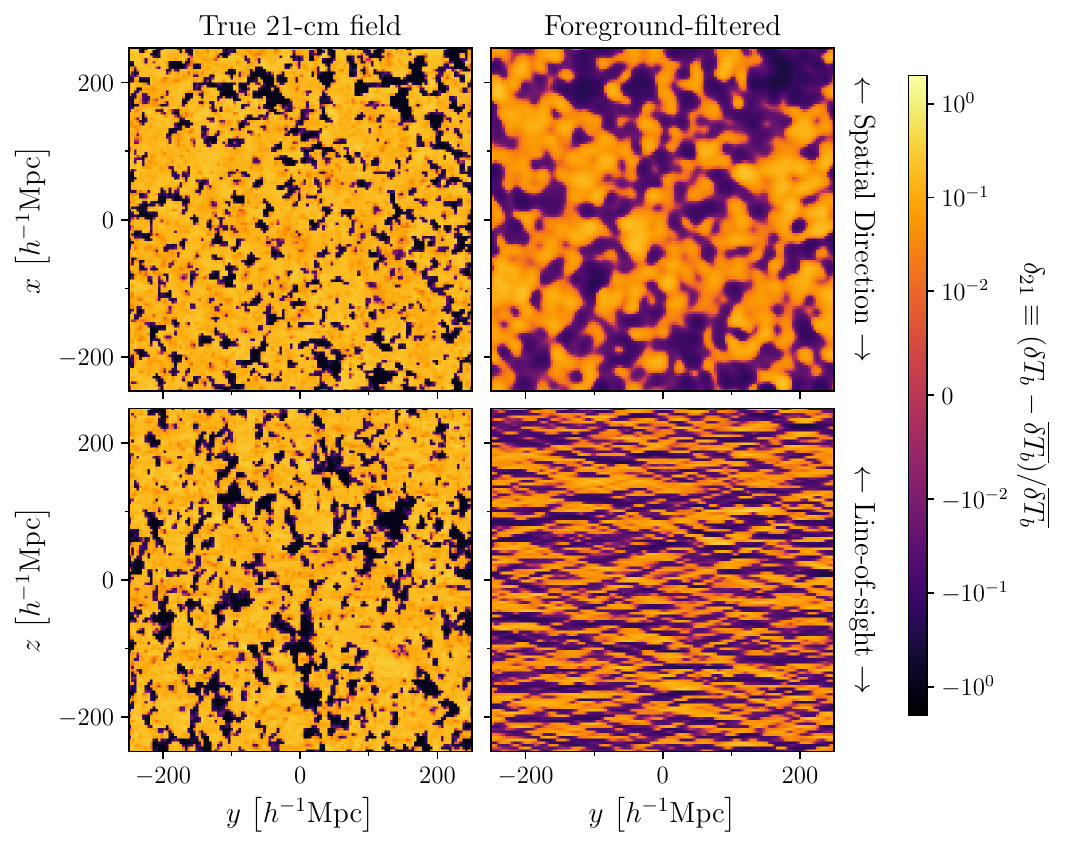}
    \caption{Examples of the true (left column) and foreground-filtered (right column) 21-cm fields generated from \texttt{21cmFAST}. The top row shows the 21-cm field in the plane perpendicular to the line of sight, while the $z$-axis in the bottom row indicates the line-of-sight direction. The foreground filter significantly distorts the line-of-sight axis in order to remove the spectrally smooth foreground contamination. 
    \label{fig:snapshot}
    }
\end{figure}

\begin{figure}[htb]
    \centering
    \includegraphics[width=\linewidth]{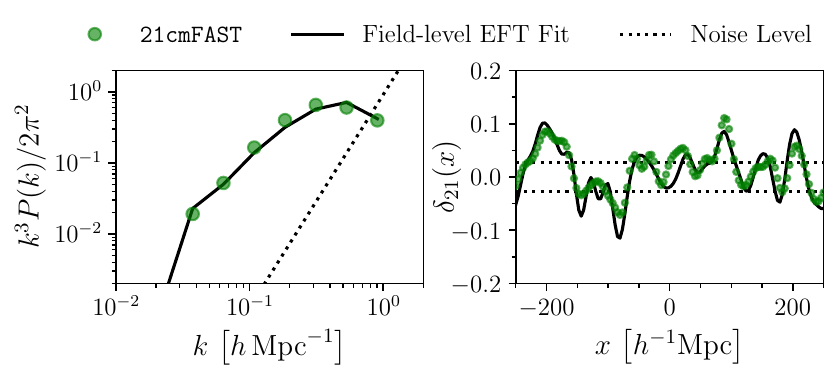}
    \caption{Foreground-filtered 21-cm signal and noise in Fourier space (left) and coordinate space (right). The left panel shows the spherically averaged power spectrum as a function of Fourier modes $k$; the right panel shows a slice of 21-cm field along a spatial coordinate. The green dots are from \texttt{21cmFAST} using the astrophysical parameters given in \autoref{tab:21cmFASTTrainingParamRange}; the solid black lines are obtained by fitting an effective field theory model to the \texttt{21cmFAST} data at the field level, given the underlying matter density field; the dotted lines indicate the noise level assumed in this work. The result on the right panel is smoothed beyond $k\sim 0.8\,h/$Mpc, a scale at which we assume the effective field theory framework becomes invalid.
    \label{fig:eft_fit_and_noise}
    }
\end{figure}

In \autoref{fig:snapshot} we show slices of the 21-cm field generated from \texttt{21cmFAST} using the fiducial values given in \autoref{tab:21cmFASTTrainingParamRange}. The top row shows the 21-cm fields in the plane perpendicular to the line of sight, while the z-axis in the bottom row indicates the line-of-sight direction. The effect of filtering out foreground using Eq.\,\eqref{eq:foreground_filter} is shown in the right column. Here, the left column is the true 21-cm field in real space and the right column shows the same field with the modes inside the $\textit{foreground wedge}$ removed. We see that this foreground filter significantly distorts the line-of-sight axis in order to remove the smooth foreground contamination. This makes a large number of cosmological modes inaccessible to observers. The goal of this work is to exploit the couplings between different modes to infer the large-scale information lost to foreground filtering. To do so, we will assume we are in the regime where we can statistically detect small-scale 21-cm fluctuations. In this work, we adopt a pixel-wise white noise such that the signal-to-noise ratio of the 21-cm power spectrum is of order one at $k\sim0.8\,h/$Mpc. This is achievable with roughly one year of full HERA observations. The noise level relative to the 21-cm signal is given as the dotted lines in \autoref{fig:eft_fit_and_noise}.

\subsection{Modeling 21-cm with Effective Field Theory} \label{subsec:eft}

If the brightness temperature field evolved linearly, it would be impossible to make any inference of the large-scale modes solely from observations of the small-scale modes. However, nonlinearities develop both through gravitational interactions and complex astrophysics. Here, we discuss the model adopted in this work to capture these non-linear mode-couplings.

While \texttt{21cmFAST} provides a natural framework to describe the observed 21-cm brightness temperature, inferences made through \texttt{21cmFAST} depend on the underlying assumptions regarding the astrophysical processes during the Cosmic Dawn and the Epoch of Reionization. Here we attempt instead to make model-independent inference by adopting the effective field theory (EFT) of 21-cm field framework developed in Refs. \cite{McQuinn:2018zwa, Qin:2022xho}. An EFT framework allows us to parametrize any bias tracer of the underlying density field with a form that is only constrained by symmetries. Following Ref. \cite{Qin:2022xho}, we define the 21-cm overdensity field as $\delta_{21}\equiv(\delta T_b-\overline{\delta T_b})/\overline{\delta T_b}$ and write

\begin{align}
\label{eq:21cm_EFT}
    \delta_{21}[\{b_i\},\delta_{\rm IC}](\kvec) &= b_{1}^{(R)}\delta(\kvec) - b_{\nabla^2}k^2\delta(\kvec) + b_2^{(R)}[\delta^2](\kvec) + b_{\mathcal{G}_2}^{(R)}[\mathcal{G}_2](\kvec) \nonumber\\
    &- i\frac{k_\|}{\mathcal{H}}\left[v_\|(\kvec) + b_1(\delta v_\|)(\kvec) - b_{\nabla^2}k^2(\delta v_\|)(\kvec)\right] - \frac{1}{2}\left(\frac{k_\|}{\mathcal{H}}\right)^2 [v_\|^2](\kvec)\,.
\end{align}

Here, $b_1^{(R)}$ is the linear bias describing the relationship between the 21-cm field and the underlying matter density field $\delta$, while $b_2^{(R)}$ accounts for nonlinearities through a quadratic bias term. The parameter $b_{\nabla^2}$ is related to the effective size of ionized bubbles with unit $[\text{Mpc}^2/h^2]$, and $b_{\mathcal{G}_2}^{(R)}$ captures local anisotropies in the field. The subscript $(R)$ denotes that a given bias parameter is $\textit{renormalized}$. In addition, $v_{\|}$ is the projection of velocity in the line-of-sight direction that describes the redshift space distortion, and $\mathcal{H}$ is the conformal Hubble parameter. The matter density field $\delta$ is obtained from the input initial condition $\delta_\mathrm{IC}$ via the third-order Lagrangian perturbation theory (3LPT). Details of 3LPT can be found in \autoref{appendix:LPT}, and we refer the reader to Ref.\,\cite{Qin:2022xho} for more details on the procedure of renormalization.

Ref.\,\cite{Qin:2022xho} demonstrated that this bias expansion accurately describes the large-scale ($k \lesssim 0.8\,h/\mathrm{Mpc}$) 21-cm field at the field level in the regime where the IGM is sufficiently heated ($T_{\rm spin} \gg T_{\rm CMB}$) and not highly ionized ($x_{\mathrm{HI}} \gtrsim 0.4$). Accordingly, in this work, we focus on results at $z = 9.0$ with a mean neutral fraction $\bar{x}_{\mathrm{HI}} = 0.64$ as a representative example. \autoref{fig:eft_fit_and_noise} shows the agreement between the EFT framework and $\texttt{21cmFAST}$. Here, we adopt a field-level fit of the 21-cm field with a known initial condition $\delta_\mathrm{IC}$. We obtain:
\begin{equation}
    \label{eq:fiducial_bias}
    (b_1^{(R)}, b_{\nabla^2}, b_2^{(R)}, b_{\mathcal{G}_2}^{(R)}) = (-1.96, -1.429, -2.177, -0.042)
\end{equation}

The EFT results generated from this set of bias parameters and a matter density field derived from 3LPT are shown as solid black lines in \autoref{fig:eft_fit_and_noise}. We can see that EFT is able to describe results from \texttt{21cmFAST} (green circles) in both summary statistics, such as power spectrum (left panel) and at the field level (right panel). Throughout this work, we adopt a scale cut of $k_{\rm cut}=0.8\,h/$Mpc to smooth all our field-level results beyond this scale\footnote{We choose the cutoff scale to be at 0.8 h/Mpc for our redshift of interest, as it was shown in Ref. \cite{Qin:2022xho} that the EFT framework can well-describe 21-cm field generated from radiation-magneto-hydrodynamic simulations up to this scale. We have further validated that our results are not significantly impacted if we adopt a slightly smaller cutoff scale.}.

\section{Field-level Inference with Effective Field Theory}\label{sec:fieldLevelInferenceWithMCLMC}

In this section, we describe how we use the EFT framework outlined in Section\,\ref{subsec:eft} to infer the missing large-scale 21-cm modes. This is achieved in two steps: (1) We use a gradient-based sampler to infer both the underlying initial condition $\delta_\mathrm{IC}$ and all the bias parameters from the observed foreground-filtered 21-cm field; (2) We calculate the 21-cm field at all scales with the inferred initial condition and bias parameters through the EFT framework allows. In Section\,\ref{subsec:implement_HMC}, we lay out the technical implementation of our method. Readers interested in the results can skip to Section\,\ref{subsec:validation_HMC}, where we show important validation tests of our method. 

\subsection{Implementation of Gradient-Based Samplers} \label{subsec:implement_HMC}

To enable computationally feasible field-level inference, we implement the EFT formalism using the Python library {\tt Jax}\footnote{\href{http://github.com/jax-ml/jax}{\color{black}http://github.com/jax-ml/jax}} \cite{jax2018github}, which provides key features such as auto-differentiation, just-in-time (JIT) compilation, and GPU acceleration. These capabilities allow us to efficiently evaluate the 21-cm field and its derivatives, which is key for gradient-based inference. With this implementation, we are able to compute one realization of the 21-cm field with a resolution of $128^3$ voxels in approximately $\mathcal{O}(100)$ milliseconds on a single NVIDIA A100 GPU. 

We then adopt the Microcanonical Langevin Monte Carlo (MCLMC) algorithm~\cite{Robnik:2022bzs,Robnik:2023pgt,Bayer:2023rmj} to sample both bias parameters and ICs simultaneously given the observed 21-cm field. Similar to the Hamiltonian Monte Carlo (HMC) algorithm, MCLMC is a gradient-based sampler that utilizes the gradient of the likelihood function to speed up the sampling process, but differs in the way it constructs the Hamiltonian. In HMC, we express the target distribution as: 
\begin{align}
    p({\color{black} y}) \propto \int_{\mathbb{R}^d}\d\Pi\, e^{-{\color{black} \mathcal{H}}({\color{black} y},\Pi)}\,,
\end{align}
where ${\color{black} y}$ is the target parameter to be sampled, $\Pi$ is the canonical momentum, and ${\color{black} \mathcal{H}}$ is the corresponding Hamiltonian defined as ${\color{black} \mathcal{H}}({\color{black} y},\Pi)=\Pi^2({\color{black} y})/2 + \mathcal{L}({\color{black} y})$. By sampling both ${\color{black} y}$ and $\Pi$ with the given Hamiltonian dynamics
\begin{align}
    \d {\color{black} y} = u \d t\,,\quad \d u = -\nabla\mathcal{L}({\color{black} y})\d t\,,
\end{align}
we could explore the posterior space more efficiently than with the traditional Metropolis-Hasting (MH) algorithm. The target distribution can then be obtained through marginalization of the canonical momentum. However, to fully explore the posterior space, the HMC algorithm has to sample different energy levels and requires an additional MH acceptance-rejection step within the algorithm. Both of these reduce the efficiency of HMC. 

The MCLMC algorithm improves over HMC according to the following. Rather than considering the marginal of the canonical distribution, MCLMC focuses on the marginal of the microcanonical distribution:
\begin{align}
    p({\color{black} y}) \propto \int_{\mathbb{R}^d}\d\Pi\,\delta_{\rm D}({\color{black} \mathcal{H}}({\color{black} y},\Pi) - E)\,,
\end{align}
where $\delta_{\rm D}(\cdot)$ denotes the delta function and $E$ is the energy. The Hamiltonian dynamics, in this case, becomes:
\begin{align}
    \d {\color{black} y} = u\d t\,,\quad \d u = P(u)\left[f({\color{black} y})\d t + \eta\d W\right],
\end{align}
with a Hamiltonian detailed in Ref.~\cite{Robnik:2023pgt}. Here, $P(u)\equiv (1-uu^T)$ is the projection matrix and $f({\color{black} y})\equiv-\nabla\mathcal{L}({\color{black} y})/(d-1)$ is the force term on the sampling particles. A diffusion term proportional to $\eta$ is introduced to improve ergodicity ($W$ is a normal random vector). In total, there are two hyperparameters: the step size and the amplitude of the diffusion term $\eta$. These hyperparameters are tuned to ensure the energy fluctuation is below a certain threshold ($5\times10^{-4}$). Therefore, we now only need to sample the momentum at a fixed energy level, which greatly improves the algorithm's efficiency over HMC. 

We define our log-likelihood function $\ln\mathcal{L}$ for the field-level inference as 
\begin{align}
    -2\ln\mathcal{L}(\{b_i\}, \delta_{\rm IC}) = \sum_{\kvec_i\notin\text{wedge}}\frac{|\delta_{\rm 21,\,obs}(\kvec_i)-\delta_{\rm 21}[\{b_i\},\,\delta_{\rm IC}](\kvec_i)|^2}{N(\kvec_i)} + \sum_{\kvec_i}\frac{|\delta_{\rm IC}(\kvec_i)|^2}{N_{\rm prior}(\kvec_i)} + C\,,
\end{align}
where $C$ is a constant, $\delta_{\rm 21,\,obs}$ denotes the observed foreground-filtered 21-cm field, $N(\kvec)$ is the noise power spectrum, and $N_{\rm prior}(\kvec)$ is the power spectrum of a density field assumed to be Gaussian distributed with unit variance in real space\footnote{See \autoref{appendix:LPT} for details on how $\delta_{\rm IC}$ is related to the matter density field within the EFT framework.}. Note that the first term only sums over 21-cm modes outside the foreground wedge, as these are the only observable ones. 

\begin{figure}
    \centering
    \includegraphics[width=1.0\linewidth]{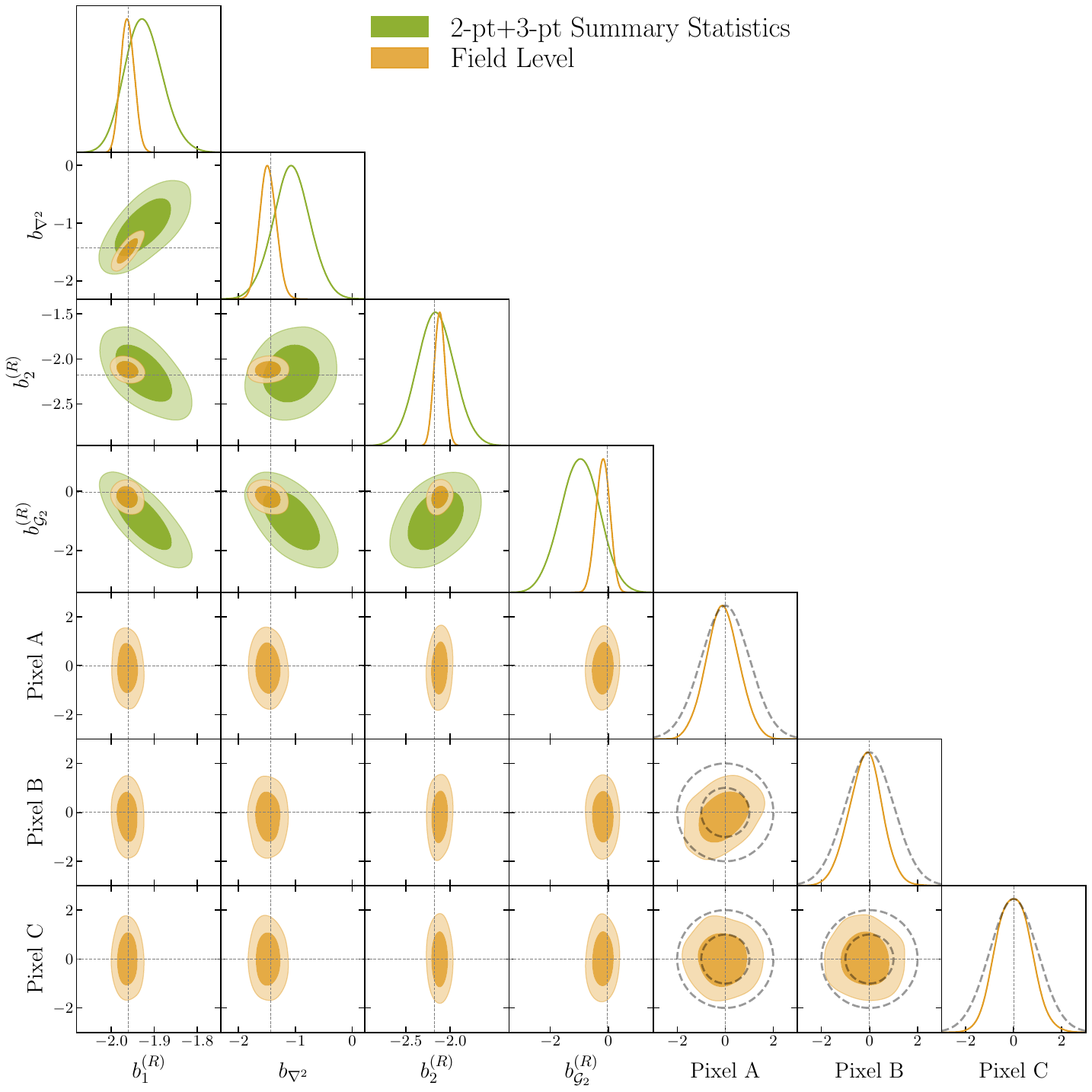}
    \caption{Renormalized bias parameters inferred from foreground-filtered, EFT-generated 21-cm field. The dashed lines show the fiducial values of the bias parameters (see Eq.\,\eqref{eq:fiducial_bias}), and the contours indicate the 68\% and 95\% confidence regions for the posterior distribution. The orange contours correspond to the field-level fit as described in Section\,\ref{subsec:implement_HMC}, while the green contours are obtained by fitting only 2-pt and 3-pt summary statistics---the power spectrum monopole and the skew spectra (c.f. Section\,\ref{subsec:summary_statistics}). As the field-level fit also constrains the initial density field, we show the posterior distributions for three pixels in real space as an example. Here, Pixel A and B are next to each other, while Pixel C lies $\sim$50 comoving $\mathrm{Mpc}/h$ away. For reference, the dashed gray curves indicate the standard normal priors assumed for each pixel. 
    \label{fig:recovery_eftbias}}
\end{figure}

To speed up the sampling, we first perform an optimization of the likelihood function with a quasi-Newtonian L-BFGS~\cite{nocedal1999numerical} algorithm to find a reasonable guess of $\delta_{\rm IC}$ using a fixed set of bias parameters estimated from the summary statistics (see the green contours in \autoref{fig:recovery_eftbias}). Following Ref.~\cite{Modi:2019hnu}, we include an iterative smoothing and an up/down-sampling scheme in our optimization pipeline to reconstruct large-scale modes first. This uses the fact that large-scale modes are more linear compared to small-scale modes, leading to a likelihood function with a more convex geometry that can increase the speed of convergence. The estimated MAP of $\delta_{\rm IC}$ and bias parameters are then used as the starting point for our MCLMC chain. We also compute the diagonal of the Hessian to rescale each parameter so that they have approximately the same variance. This preconditioning step improves numerical stability and efficiency by ensuring that all parameters are updated on comparable scales. We have checked that the results are robust under different starting points, albeit with a much lower efficiency.

We implement MCLMC with {\tt Blackjax}\footnote{\url{https://blackjax-devs.github.io/blackjax/}}, a {\tt Jax}-based library for samplers, so that a single step takes only $\mathcal{O}(100\text{ ms})$. To determine whether the sampling has converged, we use the effective sample size (ESS) to assess how effectively the samples approximate the posterior distribution. This is defined as ${\rm ESS}\equiv N/(1+2\sum_{t=1}^{\infty}\rho_t)$ where $\rho_t$ is the autocorrelation at lag $t$ and $N$ is the number of sampling steps. We ensure that $\textrm{ESS}\gtrsim200$ for most parameters.

\subsection{Summary Statistics as a Baseline for Field-Level Inference} \label{subsec:summary_statistics}
To evaluate the information gain from field-level density reconstruction, we compare against widely used summary statistics. Specifically, we consider the power spectrum monopole $P_0(k)$ and the skew-spectrum $\tilde{P}_{\mathcal{S}_i}(k)$ as our baseline. The skew-spectrum serves as a compressed statistic for bispectrum information and has been shown to retain sensitivity to amplitude-like parameters---including bias parameters---under the assumption of optimal kernel choices~\cite{Schmittfull:2014tca,MoradinezhadDizgah:2019xun}. In this work, we adopt the first three Legendre polynomials as the kernel basis, as they provide a minimal yet expressive decomposition of the angular dependence~\cite{Chen:2024bdg}. Throughout this analysis, we assume a Gaussian likelihood for the summary statistics.

We fit both statistics within a wavenumber range of $[k_{\rm min}, k_{\rm max}]$. Here, we choose $k_{\rm min} = 1.5\, k_{\rm fund}$ and $k_{\rm max} = k_{\rm cut}$, where $k_{\rm fund} = 2\pi / L_{\rm box}$ is the fundamental mode set by the box size $L_{\rm box}$, and $k_{\rm cut}$ is the same maximum wavenumber employed in the EFT-based field-level analysis. This range between $[0.02,  0.8]\,h/\mathrm{Mpc}$ is further divided into 20 linearly spaced $k$-bins for evaluating the summary statistics. 

To incorporate observational effects such as the foreground wedge and instrumental noise into the likelihood, we estimate the covariance matrix from 10{,}000 Monte Carlo realizations of the 21-cm field. Each realization is generated with independent initial conditions but fixed bias parameters, as described in Section~\ref{subsec:eft}. For the theoretical prediction at each sampling step, we forward-model the full field---including wedge filtering and noise---and compute the summary statistics accordingly. To suppress the sample variance due to stochasticity in the forward modeling, we average over $n_{\rm realization}$ (in practice, 15 realizations at each step). The resulting sampling noise introduces an additional correction to the covariance, which we account for by rescaling the covariance matrix by a factor $1 + 1/n_{\rm realization}$, assuming Gaussian-distributed fluctuations.

We will denote the combination of power spectrum monopole and skew-spectra with three Legendre polynomials as 2-pt+3-pt in our later comparison. We do not perform analysis with power spectrum monopole alone because without it, higher-order bias parameters besides $b_1^{(R)}$ cannot be well-constrained.

\subsection{Validation} \label{subsec:validation_HMC}

\begin{figure}
    \centering
    \includegraphics[width=\linewidth]{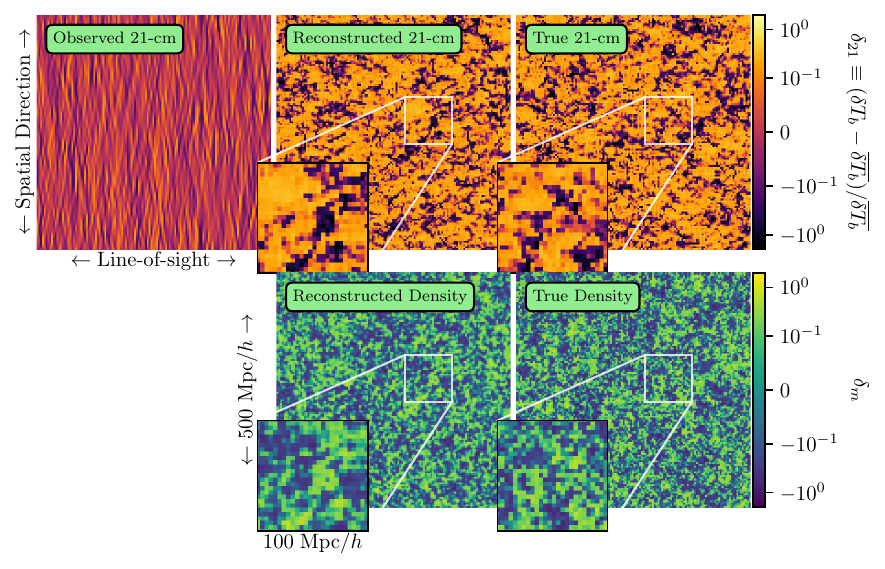}
    \caption{Configuration-space reconstructions using the gradient-based method on the foreground-filtered, EFT-generated 21-cm field. The top left corner shows a slice of the input 21-cm field with modes within the foreground wedge removed. Our gradient-based sampler is applied directly to this input to sample both the bias parameters (see \autoref{fig:recovery_eftbias}) and the initial density field. The bottom left panel shows the reconstructed density field, marginalized over bias parameters. Each joint sample of the bias parameters and the initial density field is then used to generate a 21-cm field, as can be seen in the middle panel on the top row. These maps are $500$ comoving $\mathrm{Mpc}/h$ on the side and we also zoom in to a $100\times100$ comoving $(\mathrm{Mpc}/h)^2$ area to see the performance of our method on small scales. 
    \label{fig:recovery_eftfield_real_space}}
\end{figure}

\begin{figure}
    \centering
    \includegraphics[width=\linewidth]{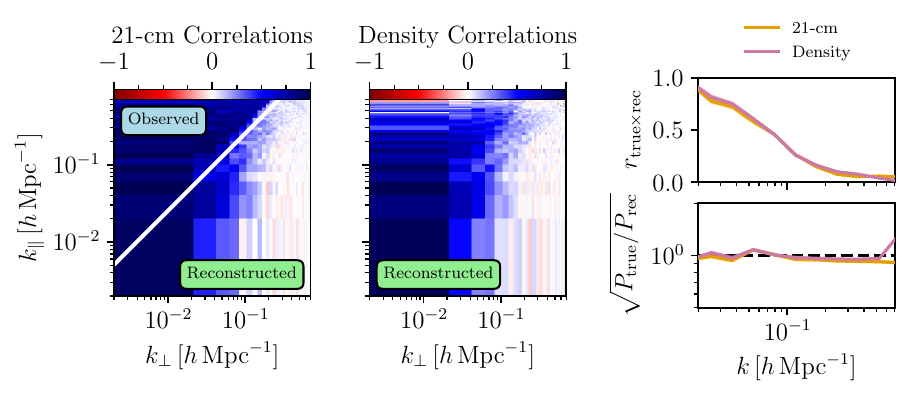}
    \caption{Validation tests of the gradient-based method on the foreground-filtered, EFT-generated 21-cm field. Left: Cross-correlation between the true underlying 21-cm field and the reconstructed 21-cm field; Middle: Cross-correlation between the true underlying initial density condition and the reconstructed one; Right: Spherically averaged cross-correlations (upper panel) and transfer functions (lower panel) between the true and reconstructed 21-cm (orange) and underlying initial density condition (purple) fields (see~\autoref{eq:recon_statistics}). Shaded regions denote the 68\% credible intervals from 100 posterior samples.
    \label{fig:recovery_eftfield}}
\end{figure}

To validate our methodology, we first perform the reconstruction on an EFT-generated 21-cm field. This ensures that our modeling framework can completely describe the input data. We generate a $(500\,\mathrm{Mpc}/h)^3$ volume with a resolution of $128^3$ voxels using the best-fit bias parameters given in Eq.\,\eqref{eq:fiducial_bias}. The EFT-generated 21-cm field is then filtered and supplemented with noise as described in Section\,\ref{subsec:eft}. 

The foreground-filtered, EFT-generated 21-cm field is then used to infer both the bias parameters and the ICs. The orange contour in \autoref{fig:recovery_eftbias} shows the posterior distribution on both the bias parameters and three real-space pixels. For comparison, we also show constraints obtained from fitting to summary statistics, specifically, the combination of the power spectrum monopole and skew-spectra with three Legendre polynomial kernels. For the summary statistics, we assume a Gaussian likelihood with the same noise level and only fit for summary statistics outside the foreground wedge as detailed in Section\,\ref{subsec:summary_statistics}. The same cutoff scale is applied to both analyses. \autoref{fig:recovery_eftbias} clearly indicates that field-level inference yields significantly tighter constraints compared to the summary statistics. Notably, the improvement is observed not only for higher-order bias parameters but also for the linear bias $b_1^{(R)}$. The fractional improvement, defined as $(\sigma_{\rm summary} - \sigma_{\rm field-level}) / \sigma_{\rm summary}$, is 64\%, 55\%, 71\%, and 63\% for $b_1^{(R)}$, $b_{\nabla^2}$, $b_2^{(R)}$, and $b_{\mathcal{G}_2}^{(R)}$, respectively.

\autoref{fig:recovery_eftbias} shows the posterior distribution for three real-space pixels as an example of the field-level constraint of the IC. We find that after posterior sampling, the standard deviations of the posteriors are reduced by approximately 30\% compared to the priors, indicating an effective constraint from the data. {\color{black}Among the three pixels, Pixel A and B are next to each other, and we observe a slight degeneracy in the posterior distribution. This is expected as we have fewer data points (the amount of Fourier modes outside the foreground wedge) than parameters (bias parameters and all modes for the ICs). Meanwhile, Pixel C is $\sim$50 Mpc$/h$ away from the other two pixels, and we see very little correlation between them. In fact, the correlation decreases dramatically as we move beyond the nearest neighbors\footnote{We note that by construction, each pixel of the ICs here should be uncorrelated with each other. The ICs are further convolved with the two-point correlation function to ensure they have the correct statistics. For detailed of this procedure, please see \autoref{appendix:LPT}.}. } Moreover, we do not find significant correlations between $b_1^{(R)}$ and the IC in real space. This is likely because small-scale modes are reconstructed less accurately, reducing their ability to introduce correlations with the bias parameters in real space. 

\autoref{fig:recovery_eftfield_real_space} shows the results of our reconstruction in configuration space. The top left corner shows a slice of the input 21-cm field in which modes within the foreground wedge have been removed. Our gradient-based sampler is applied directly to this input (plus noise), the initial density field, and bias parameters, simultaneously. We stress that the only assumption we have made here is that we know exactly the linear matter power spectrum. We do not make any additional assumption of the underlying matter density at the field level. The left panel on the bottom row shows the reconstructed density field, marginalized over bias parameters. The initial density field is then combined with the bias parameters to generate the 21-cm field using Eq.\,\eqref{eq:21cm_EFT} (middle panel of the top row). Compared to the truth on the rightmost column, we see that our gradient-based sampler is able to reconstruct large-scale information for both the 21-cm field and the initial density field. Detailed morphology of the reionization bubbles is slightly different between the reconstructed field and the truth, indicating the challenge to reconstruct small-scale modes that are more non-linear. 

To evaluate the performance on the inferred ICs and the reconstructed 21-cm field more quantitatively, we consider two metrics:

\begin{equation}
\begin{aligned}
    \textrm{Transfer Function:}\quad & \sqrt{\frac{P_\mathrm{true}(\kvec)}{P_\mathrm{rec}(\kvec)}}  \\
    & \\
    \textrm{Cross-Correlation:}\quad & r_{\textrm{true}\times\textrm{rec}} \equiv \textrm{Real}\left[\frac{\delta^*_\mathrm{true}(\kvec)\times\delta_\mathrm{rec}(\kvec)}{\sqrt{P_\mathrm{true}(\kvec)P_\mathrm{rec}(\kvec)}}\right],
\end{aligned}\label{eq:recon_statistics}
\end{equation}
which measure the amplitude and the phase of the reconstructed fields, respectively. \autoref{fig:recovery_eftfield} shows these metrics for our validation tests. The left and middle panels show the cross-correlation coefficient $r_{\textrm{true}\times\textrm{rec}}$ on the two-dimensional $k$-space for both the inferred initial condition and the reconstructed 21-cm field. The white line in the leftmost panel indicates where the \textit{foreground wedge} is located. We see that the ``reconstructed'' 21-cm field outside the foreground wedge shows great agreement with the ground truth, indicating that our framework is self-consistent. Meanwhile, both the 21-cm modes inside the foreground wedge and the initial condition across Fourier space are not known to us \textit{a priori}. Correlations in these regimes indicate that our method is highly successful. The right two panels of \autoref{fig:recovery_eftfield} show the spherically averaged cross-correlation and transfer function. In the top panel, we see that the correlation decreases at small scales. As we recover most modes outside the foreground wedge, these spherically averaged correlations are a measure of the performance of our reconstruction for modes within the wedge. {\color{black} At $k=0.02\,h$/Mpc, we are able to recover $89\%$ of the modes. The performance drops at smaller scales, with $35\%$ ($12\%$) of modes recovered at $k=0.1~(0.2)\,h/$Mpc. This reflects the combined impact of instrumental noise, aggressive foreground filtering, and non‐linear mode–coupling that increasingly degrades phase information on small scales.} Lastly, the bottom right panel shows the transfer function for both the ICs and the 21-cm field. Here, our method over-estimates the power of the ICs and the reconstructed 21-cm field by at most $5\%$ {\color{black} at large scales, gradually increasing to 10\% at smaller scales after $k=0.1\,h/$Mpc for the 21-cm field}. At large scales, the amplitude of the power spectrum is mainly determined by $b_1$, but terms proportional to $b_{\nabla^2}$ start to contribute more at larger $k$. Uncertainties in these terms likely induce the deviation we observe at smaller scales.

It is important to emphasize that the constraints presented here represent a conservative estimate of the potential information gain from non-Gaussian statistics. Due to the perturbative nature of the EFT-based forward model, we have to limit ourselves to the highest $k$ modes allowed within the theory. Those discarded small-scale modes may further enhance the constraining power of field-level inference, if modeled correctly. Despite this conservative scale cut, we still observe significant improvements over the summary-statistics-based approach, highlighting the strength of non-Gaussian information during reionization. Moreover, these results also suggest that field-level methods provide a powerful framework not only for bias parameter estimation but also for cosmological parameter inference from 21-cm observations.

\section{Inference with Diffusion-Based Generative Model}\label{sec:diffusionBasedGenerativeModel}

In this section, we present another approach for recovering the wedge-suppressed modes of the 21-cm signal using a diffusion-based generative model. This method involves two stages as well: (1) We train a conditional generative model to learn the mapping between observed wedge-filtered fields and the true underlying 21-cm field; (2) We generate posterior samples through an iterative denoising process known as ancestral sampling. Section~\ref{subsec:variational_diffusion_model} outlines the theoretical framework and model architecture. Readers interested in the results can skip to Section~\ref{subsec:validation_diffusion} where we show important validation tests of our method. 

\subsection{Variational Diffusion Model}\label{subsec:variational_diffusion_model}

The diffusion model has demonstrated stability and scalability for image generation, outperforming methods such as normalizing flows and generative adversarial networks (GANs). Reference~\cite{2021arXiv210700630K} establishes the probabilistic foundation of this approach, showing that minimizing the loss function is equivalent to approximating the true sampling process of the posterior distribution. 

In our framework, we construct the generative model $p_\btheta(\x|\c)$ conditioned on $\c$ for a given sample $\x$ by reversing the diffusion process starting from a Gaussian random field. The corresponding log-probability function is expressed as: 
\begin{align}
\log p_\btheta(\x|\c) = \log \int_{\{{\color{black}\y}_{i}:\,i=1,\dots,T\}} p_\btheta({\color{black}\y}_1|\c) p_\btheta(\x|{\color{black}\y}_0, \c) \prod_{i=1}^{T} p_\btheta({\color{black}\y}_{s(i)}|{\color{black}\y}_{t(i)}, \c)\,,
\end{align}
where $\btheta$ represents the trainable parameters of the diffusion model, $T$ is the number of time steps, and ${\color{black}\y}_i$ are latent variables generated by the Gaussian diffusion process. Here, the time steps follow $1 > t(i) = i/T > s(i) = (i-1)/T > 0$, and by construction, $p_\btheta({\color{black}\y}_1|\c) \sim \mathcal{N}(0,\mathbf{I})$. To generate a posterior sample, we start with a Gaussian random field sampled from $\mathcal{N}(0,\mathbf{I})$ and iteratively denoise it using the learned model $p_\btheta({\color{black}\y}_{s(i)}|{\color{black}\y}_{t(i)}, \c)$. This iterative denoising process follows a stochastic differential equation and continues until the final step, yielding a sample from the posterior distribution.

For the training of the diffusion model, we first generate noisy versions/latent variables of the data through a forward Gaussian diffusion process with a noise schedule
\begin{align}
    \sigma_t^2 = \text{sigmoid}(\gamma_{\boldsymbol{\eta}}(t))\,,
\end{align}
where $\gamma_{\boldsymbol{\eta}}(t)$ is a learnable linear function used to determine the noise level added at each single time step. We also consider a variance-preserving map during the forward process such that the variance of the field limits the field values within a certain range. As a result the PDF of ${\color{black}\y}_t$ given $\x$ at time $t$ can be written as $q({\color{black}\y}_t|\x) = \mathcal{N}(\alpha_t \x,\sigma_t^2\mathbf{I})$ with $\alpha_t=\sqrt{1-\sigma_t^2}$. Next, we train the model to revert/denoise the above diffusion process at each time step by minimizing the following variational lower bound $\text{VLB}(\x)$ to approximate the true likelihood function:
\begin{align}
    -\log p_\btheta(\x|\c) \leq \text{VLB}(\x) = D_{\rm KL}(q({\color{black}\y}_1|\x,\c)||p_\theta({\color{black}\y}_1|\c))+\mathbb{E}_{q({\color{black}\y}_0|\x,\c)}[-\log p_\theta(\x|{\color{black}\y}_0,\c)] + \mathcal{L}_{T}(\x|\c)\,,
\end{align}
where VLB$(\x)\equiv\mathbb{E}_{{\color{black}\y}\sim q(\cdot|\x,\c)}\left[\log(p_\btheta(\x,\boldsymbol{{\color{black}\y}}|\c)/q({\color{black}\y}|\x,\c))\right]$, $D_{\rm KL}$ is the {\color{black} Kullback–Leibler (KL)-divergence} used to measure the distance between two PDFs and $\mathcal{L}_T(\x|\c)$ is the diffusion loss with discrete time step given by:
\begin{align}
    \mathcal{L}_T(\x|\c)&=\sum_{i=1}^T\mathbb{E}_{q({\color{black}\y}_{t(i)}|\x)}D_{\rm KL}\left[q({\color{black}\y}_{s(i)}|{\color{black}\y}_{t(i)},\x,\c)||p_\btheta({\color{black}\y}_{s(i)}|{\color{black}\y}_{t(i)},\c)\right]\nonumber\\
    &=\frac{T}{2}\mathbb{E}_{\bepsilon\sim\mathcal{N}(0,\mathbf{I}),i\sim U_{[1,T]}}\left[(\exp(\gamma_{\boldsymbol{\eta}}(t)-\gamma_{\boldsymbol{\eta}}(s))-1)||\bepsilon-\hat{\bepsilon}_\theta({\color{black}\y}_t,t,\c)||_2^2\right],
\end{align}
with $\hat{\bepsilon}_\btheta$ modeled by the neural net. Once we minimize the variational lower bound to obtain the approximate likelihood function, the posterior sampling can be done with the ancestral sampling:
\begin{align}
    {\color{black}\y}_s = \sqrt{\frac{\alpha_s^2}{\alpha_t^2}}\left({\color{black}\y}_t-\sigma_t a_{\boldsymbol{\eta}}(s,t)\hat{\bepsilon}_\theta({\color{black}\y}_t,t,\c)\right) + \sqrt{(1-\alpha_s^2)a_{\boldsymbol{\eta}}(s,t)}\,\bepsilon
\end{align}
where $a_{\boldsymbol{\eta}}(s,t)=1-\exp(\gamma_{\boldsymbol{\eta}}(s)-\gamma_{\boldsymbol{\eta}}(t))$ and $\bepsilon\sim\mathcal{N}(0,\mathbf{I})$. In this work, we consider 150 time steps for our posterior sampling, and we find almost no improvements when we increase it.

To apply the diffusion model to our case for wedge recovery, $\x$ becomes the true 21-cm field and the condition $\c$ is the observed foreground-filtered 21-cm field with the inclusion of noise. We train our model on two mock data sets: one generated with EFT and one with {\tt 21cmFAST}. We follow the architecture used in Refs.~\cite{Ono:2024jhn,Park:2023ync}. We use a ResUNet~\cite{2015arXiv150504597R,2016cvpr.confE...1H} as our backbone for $\hat{\epsilon}_\theta$, which involves four blocks of three-dimensional convolution followed by strided downsampling layers. Group normalization~\cite{2018arXiv180308494W} is included. We employ a learnable linear noise schedule initialized as $\gamma_{\boldsymbol{\eta}}(t)=\eta_1 t-\eta_2=26.6t-13.3$. For the optimizer, we apply AdamW~\cite{2017arXiv171105101L} with a learning rate of $1\times10^{-4}$ together with a CosineAnnealingWarmRestarts~\cite{2016arXiv160803983L} scheduler that could lower the learning rate down to $1\times10^{-5}$ periodically for both exploration and fine-tuning.

\subsection{Training Procedure}\label{subsec:training_procedure}

First, in a warm-up phase, we train a baseline model on a set of simulations on a [500 Mpc$/h$]$^3$ box with $128^3$ voxels, generated by a Sobol sequence of bias parameters, for 195,000 steps, or 285 epochs. This consists of $16,384$ simulations and the range of bias parameters are:
\begin{align}
    b_1\in[-3,\,-0.5],\quad b_{\nabla^2}\in[-3,\,-0.5],\quad b_2\in[-3,\,-0.5],\quad b_{\mathcal{G}_2}\in[-1,1]\,.
\end{align}
We include random shifting, rotation, and voxel-wise noise for data augmentation. Our batch size is chosen to be 8, and we accumulate the gradient every two steps, so the effective batch size is 16. When we train our model, we also implement data parallelization to distribute the training onto two 80GB NVIDIA A100 GPUs simultaneously, enabling a larger batch size. This model is then used as the starting point for later studies.

Next, we further fine-tune our model for 120,000 additional steps using either EFT-generated data or {\tt 21cmFAST}-generated data. For the EFT-generated data, we train our model further on a much narrower range of prior determined by the posterior distribution provided by the summary statistics, that is, posterior mean $\pm$ 3$\times $posterior standard deviation from the summary statistic. On the other hand, for the case of {\tt 21cmFAST}-generated data, the training set is generated using the parameter range shown in Table~\ref{tab:21cmFASTTrainingParamRange} at the 68\% credible level. These two training sets are generated with a Sobol sequence of $65,536$ samples for either the bias or astrophysical parameters. The choice of 65,536 simulations was primarily driven by computational resource constraints. While this number provides reasonably good coverage of the parameter space and leads to stable posterior reconstructions across test cases, we expect that increasing the training set size may further improve performance. Exploring this direction is a natural next step as additional resources become available. 

We fix the cosmological parameters to Planck 2018 best-fit values as these are better measured by other experiments currently. For the validation set, we generate $1024$ samples from uniform distribution with bounds set to be the same as the training set.

\subsection{Validation}
\label{subsec:validation_diffusion}
\begin{figure}
    \centering
    \includegraphics[width=0.8\linewidth]{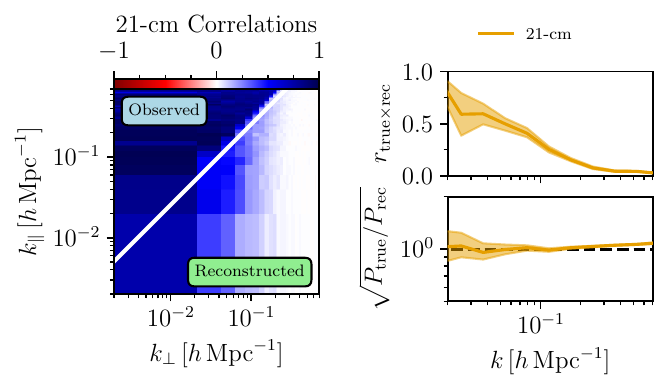}
    \caption{Validation tests of the diffusion-based method on foreground-filtered, EFT-generated 21-cm field. Left: Cross-correlation between the true underlying 21-cm field and the reconstructed 21-cm field; Right: Spherically averaged cross-correlation (upper panel) and transfer function (lower panel) between the true and reconstructed 21-cm (orange) and underlying density (purple) fields. Compared to \autoref{fig:recovery_eftfield}, the diffusion model here reconstructs the 21-cm field directly, skipping the initial density reconstruction. Shaded regions denote the 68\% credible intervals from 90 posterior samples.}
    \label{fig:recovery_eftfield_diffusion}
\end{figure}

Similar to Section~\ref{sec:fieldLevelInferenceWithMCLMC}, we perform a validation test on the EFT-generated mock. As detailed in Section~\ref{subsec:training_procedure}, the diffusion model is trained on EFT-generated mocks with the inclusion of a foreground wedge and white noise. \autoref{fig:recovery_eftfield_diffusion} shows the result when the diffusion model is both trained and tested on the EFT-generated mock. Unlike the previous one, we only show the cross-correlation and the transfer function between the true and the reconstructed 21-cm field here because the diffusion model reconstructs the 21-cm field directly without sampling the initial condition and doing the whole forward modeling. We can see, from the figure, that the performance of the reconstruction is almost identical to the field-level inference, showing that the diffusion model can learn the generative process really well. However, we also observe larger error bars for both statistics. This might be due to the fact that the diffusion model does not know how modes are correlated in the forward model beforehand, so it will need to learn this directly from the training dataset, together with the Gaussian likelihood at each voxel in Fourier space.

\section{Results \& Discussion}\label{sec:comparisonDiscussions}

We have now seen that both the gradient-based sampler and the diffusion model can successfully reconstruct large-scale 21-cm modes given a foreground-filtered, EFT-generated input data. This ensures that both methods work in the regime where our model can perfectly describe the input observations. This might not be the case in reality. In this section, we discuss and compare the performance of these two methods by applying them to more realistic simulations generated with \texttt{21cmFAST}.

For the gradient-based sampler method introduced in Section\,\ref{sec:fieldLevelInferenceWithMCLMC}, this introduces an extra modeling uncertainty. In \autoref{fig:eft_fit_and_noise}, we show the results of fitting the EFT framework to the \texttt{21cmFAST} simulation with a fixed initial condition. In this scenario, we observe a $\sim 10\%$ discrepancy between EFT and \texttt{21cmFAST} at smaller scales, giving us an estimate of the modeling uncertainties. For the diffusion model introduced in Section\,\ref{sec:diffusionBasedGenerativeModel}, the uncertainties depend on the training dataset. For diffusion models trained on EFT data, a similar $\sim 10\%$ of modeling uncertainties is expected. On the other hand, we can also train diffusion models on \texttt{21cmFAST}-generated data. In this scenario, the uncertainties come from the unknown astrophysical parameters. 

\begin{figure}
    \centering
    \includegraphics[width=0.7\linewidth]{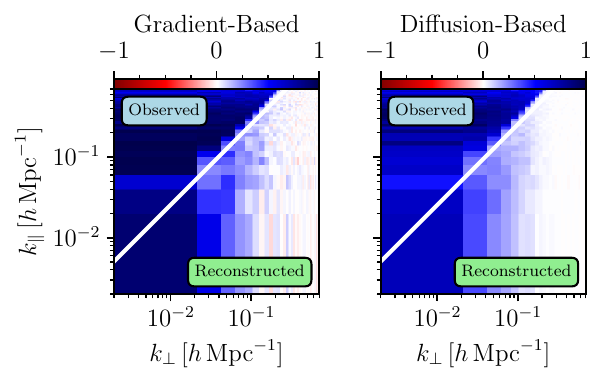}
    \caption{Results of gradient-based and diffusion-based reconstructions on foreground-filtered {\tt 21cmFAST}-generated 21-cm fields. Left: Cross-correlation between the true and the reconstructed 21-cm field using the gradient-based method with EFT modeling. Right: Same as the left panel, but with the diffusion-based reconstruction.}
    \label{fig:comparison_approaches_2d}
\end{figure}

Before comparing the two methods quantitatively, we begin with a qualitative visual inspection. \autoref{fig:comparison_approaches_2d} displays 2D reconstructions from the gradient-based sampler and the diffusion model, respectively, when applied to a \texttt{21cmFAST}-generated mock. Both methods successfully recover modes within the wedge and exhibit performance comparable to previous full Bayesian approaches. We can see that they both have the same performance on mode reconstruction within the wedge-shaped region up to a certain scale. 

\begin{figure}
    \centering
    \includegraphics[width=\linewidth]{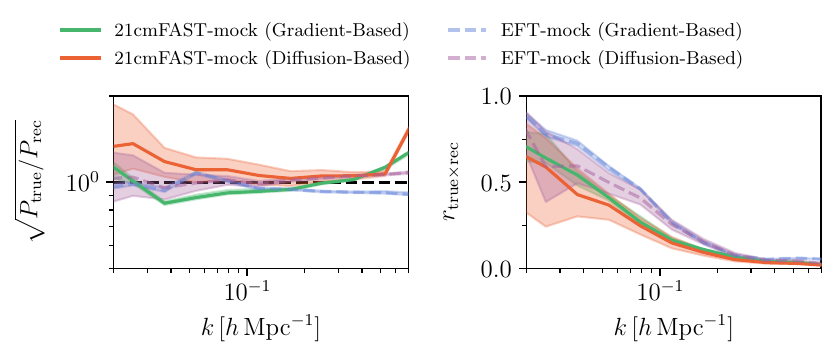}
    \caption{Estimated transfer functions (left) and cross-correlations with the truth (right) under different approaches. The solid lines show gradient-based (green) and diffusion-based (orange) method applied to {\tt 21cmFAST}-generated mocks, while the blue and purple dashed lines show the corresponding results with EFT-generated mocks. Shaded regions indicate the $68\%$ credible intervals across realizations.}
    \label{fig:comparison_approaches}
\end{figure}

A more detailed comparison is shown in \autoref{fig:comparison_approaches}. When applied to EFT-generated mocks, the field-level inference method using MCLMC outperforms the diffusion model at large scales, although their results start to converge at $k \sim 0.1\,h/\mathrm{Mpc}$ {\color{black} for the cross-correlations and at $k\sim0.2\,h$/Mpc for the transfer function}. This is expected, as the EFT mock represents an in-distribution test case for both methods. Moreover, the MCLMC sampler leverages the full likelihood and forward model explicitly, while the diffusion model must learn these from data. Still, we find that the diffusion model’s confidence intervals encompass the MCLMC results, suggesting a reasonable agreement between the two approaches.

On the more realistic \texttt{21cmFAST}-generated mock, field-level inference with MCLMC is technically an out-of-distribution application. Nevertheless, it continues to slightly outperform the diffusion model—even one trained directly on \texttt{21cmFAST} simulations—though the performance gap is narrower than in the EFT case. This suggests that while the diffusion model can approximate the underlying likelihood or data-generating process to some extent, there remains room for improvement. Ideally, it should perform better than the Bayesian approach when trained on data from the same generative process, as it is an in-distribution test. 

\begin{figure}
    \centering
    \includegraphics[width=\linewidth]{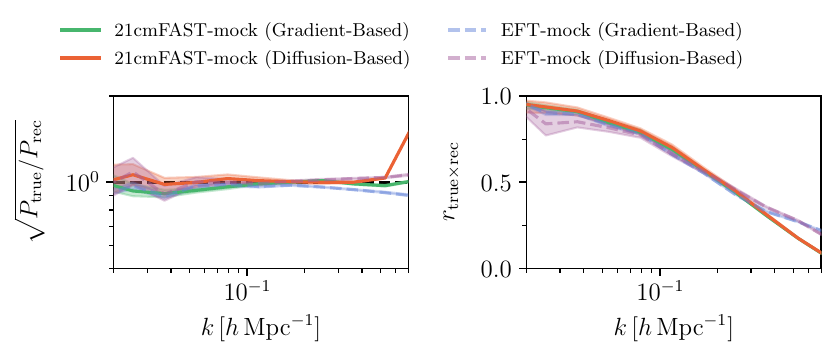}
    \caption{Similar to \autoref{fig:comparison_approaches}, but with a four-time smaller foreground wedge compared to our baseline study. This means an increased number of observed modes and a dramatic increase in performance across all our approaches. Such a scenario could be possible with the advancement of foreground subtraction methods or with carefully designed instruments that minimize the foreground contamination. }
    \label{fig:comparison_approaches_lesswedge}
\end{figure}

Finally, we have conducted an additional experiment using a significantly less aggressive wedge, where the slope is reduced by a factor of four. This scenario could be possible with the advancement of foreground subtraction methods  \cite{Datta2010:FGwedge, Liu2011:QE, Chapman2012:FastICA, Ghosh2015:Bayesian_21cm, Ewall-Wice2021:DAYNENU, Mertens2018:GPR, Kern2021:GPR, Chege2022:MWA_foreground, Wang2024:Foreground_subtraction, Mertens2024:GPR} or with carefully designed instruments that minimize foreground contamination \cite{MacKay2025:RULES}. The result is shown in \autoref{fig:comparison_approaches_lesswedge}. This setup allows us to further probe the robustness and sensitivity of both methods under varying foreground filtering conditions. In this experiment, we find that the diffusion model can slightly outperform MCLMC with EFT for the {\tt 21cmFAST}-generated mocks, provided that the model is properly trained. This result suggests that, with appropriate training, the diffusion model is capable of adapting to different foreground treatments and may offer improved reconstruction performance under a range of observational conditions. {\color{black}We note that this result is particularly relevant for experiments that subtract the foreground or adopt a hybrid foreground mitigation approach. Currently, the Low-Frequency Array (LOFAR) has been setting 21-cm limits at $k\sim0.08\,h/\mathrm{Mpc}$ \cite{Mertens2025:LOFAR_limits, Ceccotti2025:LOFAR_limit} while the Murchison Widefield Array (MWA) has been setting limits at $k\sim0.1\rm{-}0.2\,h/\mathrm{Mpc}$ \cite{Trott2020:MWA_Limit, Nunhokee2025:MWA_limit}. For comparison, experiments such as HERA, which completely avoid the foregrounds, typically set limits at $k\sim0.3\rm{-}0.5\,h/\mathrm{Mpc}$ \cite{HERA2022:h1c_idr2_limit, HERA2023:h1c_idr3_limit, HERA2025:PhaseII_Limit}} 

\section{Conclusion}
In this work, we explored two complementary approaches for reconstructing large-scale 21-cm modes lost to foreground contamination in radio interferometric observations. First, we implemented a field-level inference framework using the effective field theory (EFT) of biased tracers and a gradient-based sampler (MCLMC) to jointly infer the initial density field and bias parameters from foreground-filtered 21-cm maps. Second, we developed a variational diffusion-based generative model trained on mock data, enabling direct sampling of reconstructed 21-cm fields conditioned on the observed maps.

We validated both methods using simulations where the data-generation process is fully known. On EFT-generated mocks, the field-level Bayesian approach outperformed the diffusion model at large scales, as expected due to its access to the exact forward model. Nevertheless, the diffusion model achieved comparable reconstruction quality and retained good phase information, demonstrating its ability to learn complex correlations in the data.

Applying both methods to more realistic {\tt 21cmFAST}-generated mocks, we found that the field-level inference method, though technically out-of-distribution, continued to perform slightly better than the diffusion model. However, we observed that with appropriate training, the diffusion model could outperform the EFT-based sampler in scenarios with modified wedge shapes. This highlights the potential of deep generative models to adapt to a wider range of foreground filtering conditions and remain robust even when physical modeling becomes more uncertain.

Looking ahead, a promising path lies in combining the strengths of both approaches. The field-level framework provides a physically grounded likelihood, while the diffusion model offers scalability to smaller scales. Hybrid strategies that integrate simulation-based inference with perturbative modeling may enhance our ability to reconstruct the full 21-cm signal and extract cosmological information from previously inaccessible modes. Our results demonstrate that large-scale mode recovery from contaminated 21-cm observations is both feasible and powerful, opening new possibilities for future high-redshift surveys.

{\color{black}These reconstruction methods will be valuable for various high-redshift line-intensity mapping surveys, where foreground contamination severely limits access to large regions of Fourier space.} Recovering the information lost to the wedge not only strengthens the standalone constraining power but also enables robust cross-correlations with other tracers \cite{Hutter2017:21cm_LAE, Heneka2021:21cm_LAE, Davies2021:21cm_stack, Trott2021:MWA_LAEs, Cox2022:cross-correlation, LaPlante2023:21cm_x_Roman, Gagnon-Hartman2025:21cm_x_LAE, Hutter2025:Cross_Correlation, Chen2025:21cm_x_LAEs}. This enhances resilience to systematics and improves sensitivity to fundamental physics, including primordial non-Gaussianity and the sum of neutrino masses. Incorporating the full observational pipeline~\cite{Kern:2025bqm} into our framework would further enable realistic reconstructions of the initial conditions from observed data.

\section{Acknowledgment}
{\color{black}We thank an anonymous referee for valuable comments and suggestions.} We thank Wenzer Qin for valuable discussions and for sharing the EFT-based code, which we used to validate our implementation. We also thank Prish Chakraborty, Carolina Cuesta-Lazaro, Adélie Gorce, Sabrina Berger, Coleman Krawczyk, Adrian Liu, Jakob Robnik, and Katelin Schutz for valuable suggestions that helped improve this work. CD and SFC were partially supported by the Department of Energy (DOE) Grant No. DE-SC0020223. K.-F.C. acknowledges support from the Mitacs Globalink Research Award, Taiwan Think Global Education Trust Scholarship, and the Taiwan Ministry of Education’s Government Scholarship to Study Abroad.

\appendix
\section{Lagrangian Perturbation Theory}\label{appendix:LPT}
In this section, we will describe the way we implement the third-order Lagrangian perturbation theory (LPT) in our EFT forward model. We follow the same approach detailed in Ref.~\cite{Michaux:2020yis}. The map from the initial Lagrangian space $\boldsymbol{q}$ to the Eulerian space $\boldsymbol{x}$ at any redshift $z$ is:
\begin{align}
    \boldsymbol{q} \rightarrow \boldsymbol{x}(\boldsymbol{q},z) = \boldsymbol{q} + \boldsymbol{\psi}(\boldsymbol{q},z)\,,
\end{align}
where $\boldsymbol{\psi}$ is the displacement field that can be expanded in terms of powers of the growth function $D(z)$. We consider an expansion up to the third order, denoted as 3LPT, which can be written as:
\begin{align}
    \boldsymbol{\psi}(\boldsymbol{q},z) \approx \boldsymbol{\psi}^{(1)}(\boldsymbol{q})D(z) + \boldsymbol{\psi}^{(2)}(\boldsymbol{q})D(z)^2 + \boldsymbol{\psi}^{(3)}(\boldsymbol{q})D(z)^3 + \cdots\,.
\end{align}
These coefficients $\boldsymbol{\psi}^{(i)}$ at each position $\boldsymbol{q}$ can be derived analytically as
\begin{align}
    \boldsymbol{\psi}^{(1)}(\boldsymbol{q}) &= -\boldsymbol{\nabla}\Phi^{(1)}(q)\,,\nonumber\\
    \boldsymbol{\psi}^{(2)}(\boldsymbol{q}) &= -\frac{3}{7}\boldsymbol{\nabla}\Phi^{(2)}(q)\,,\nonumber\\
    \boldsymbol{\psi}^{(3)}(\boldsymbol{q}) &= \frac{1}{3}\boldsymbol{\nabla}\Phi^{(3a)} - \frac{10}{21}\boldsymbol{\nabla}\Phi^{(3b)} + \frac{1}{7}\boldsymbol{\nabla}\times \boldsymbol{A}^{(3c)}\,,
\end{align}
where $\Phi^{(1)}$ is the gravitational potential of the field as $a\rightarrow0$ and
\begin{alignat}{2}
    \Phi^{(2)} &= \frac{1}{2}\nabla^{-2}\left[\Phi_{,ii}^{(1)}\Phi_{,jj}^{(1)}-\Phi_{,ij}^{(1)}\Phi_{,ij}^{(1)}\right]\,,\quad\Phi^{(3a)} &&= \nabla^{-2}\left[\text{det}\Phi_{,ij}^{(1)}\right]\,,\nonumber\\
    \Phi^{(3b)} &= \frac{1}{2}\nabla^{-2}\left[\Phi_{,ii}^{(2)}\Phi_{,jj}^{(1)}-\Phi_{,ij}^{(2)}\Phi_{,ij}^{(1)}\right]\,,\quad \boldsymbol{A}^{(3c)} &&= \nabla^{-2}\left[\boldsymbol{\nabla}\Phi_{,i}^{(2)}\times\boldsymbol{\nabla}\Phi_{,i}^{(1)}\right]\,.
\end{alignat}
In particular, $\Phi^{(1)}$ is obtained by first generating a Gaussian random field drawn from a standard normal distribution, $\delta_{\rm IC}$, then multiplying by the square root of the matter power spectrum $P_{mm}(k)$, and finally applying the inverse Laplacian to account for the Poisson equation relating the gravitational potential to the density field. That is, $\Phi^{(1)}(\boldsymbol{k})=\nabla^{-2}[\sqrt{P_{mm}(k)}\,\delta_{\rm IC}(\boldsymbol{k})]$.

Once we obtain these displacement fields, we use them to displace the particles that are uniformly scattered in the simulation box and use the Cloud-In-Cell (CIC) interpolation scheme to paint them onto the grids. Compensation is implemented simultaneously to correct for the alias effect arising from the finite grid size~\cite{Jing:2004fq}. By default, the number of particles is set to be three times the number of voxels we have for the simulation.

\newpage
\bibliographystyle{utphys}
\bibliography{ref}

\providecommand{\href}[2]{#2}\begingroup\raggedright\begin{thebibliography}{100}

\bibitem{Furlanetto2006:Review}
S.~R. {Furlanetto}, S.~P. {Oh}, and F.~H. {Briggs}, ``{Cosmology at low frequencies: The 21 cm transition and the high-redshift Universe},'' \href{http://dx.doi.org/10.1016/j.physrep.2006.08.002}{{\em \physrep} {\bfseries 433} no.~4-6, (Oct., 2006) 181--301}, \href{http://arxiv.org/abs/astro-ph/0608032}{{\ttfamily arXiv:astro-ph/0608032 [astro-ph]}}.

\bibitem{Pritchard2012:Review}
J.~R. {Pritchard} and A.~{Loeb}, ``{21 cm cosmology in the 21st century},'' \href{http://dx.doi.org/10.1088/0034-4885/75/8/086901}{{\em Reports on Progress in Physics} {\bfseries 75} no.~8, (Aug., 2012) 086901}, \href{http://arxiv.org/abs/1109.6012}{{\ttfamily arXiv:1109.6012 [astro-ph.CO]}}.

\bibitem{CHIME2022:Overview}
{CHIME Collaboration}, M.~{Amiri}, K.~{Bandura}, A.~{Boskovic}, T.~{Chen}, J.-F. {Cliche}, M.~{Deng}, N.~{Denman}, M.~{Dobbs}, M.~{Fandino}, S.~{Foreman}, M.~{Halpern}, D.~{Hanna}, A.~S. {Hill}, G.~{Hinshaw}, C.~{H{\"o}fer}, J.~{Kania}, P.~{Klages}, T.~L. {Landecker}, J.~{MacEachern}, K.~{Masui}, J.~{Mena-Parra}, N.~{Milutinovic}, A.~{Mirhosseini}, L.~{Newburgh}, R.~{Nitsche}, A.~{Ordog}, U.-L. {Pen}, T.~{Pinsonneault-Marotte}, A.~{Polzin}, A.~{Reda}, A.~{Renard}, J.~R. {Shaw}, S.~R. {Siegel}, S.~{Singh}, R.~{Smegal}, I.~{Tretyakov}, K.~{van Gassen}, K.~{Vanderlinde}, H.~{Wang}, D.~V. {Wiebe}, J.~S. {Willis}, and D.~{Wulf}, ``{An Overview of CHIME, the Canadian Hydrogen Intensity Mapping Experiment},'' \href{http://dx.doi.org/10.3847/1538-4365/ac6fd9}{{\em \apjs} {\bfseries 261} no.~2, (Aug., 2022) 29}, \href{http://arxiv.org/abs/2201.07869}{{\ttfamily arXiv:2201.07869 [astro-ph.IM]}}.

\bibitem{MeerKAT2016:MeerKLASS}
M.~{Santos}, P.~{Bull}, S.~{Camera}, S.~{Chen}, J.~{Fonseca}, I.~{Heywood}, M.~{Hilton}, M.~{Jarvis}, G.~I.~G. {Jozsa}, K.~{Knowles}, L.~{Leeuw}, R.~{Maartens}, E.~{Malefahlo}, K.~{McAlpine}, K.~{Moodley}, P.~{Patel}, A.~{Pourtsidou}, M.~{Prescott}, K.~{Spekkens}, R.~{Taylor}, A.~{Witzemann}, and I.~H. {Whittam}, \href{http://dx.doi.org/10.22323/1.277.0032}{``{A Large Sky Survey with MeerKAT},''} in {\em MeerKAT Science: On the Pathway to the SKA}, p.~32.
\newblock Jan., 2016.
\newblock \href{http://arxiv.org/abs/1709.06099}{{\ttfamily arXiv:1709.06099 [astro-ph.CO]}}.

\bibitem{CHORD2019:Overview}
K.~{Vanderlinde}, A.~{Liu}, B.~{Gaensler}, D.~{Bond}, G.~{Hinshaw}, C.~{Ng}, C.~{Chiang}, I.~{Stairs}, J.-A. {Brown}, J.~{Sievers}, J.~{Mena}, K.~{Smith}, K.~{Bandura}, K.~{Masui}, K.~{Spekkens}, L.~{Belostotski}, M.~{Dobbs}, N.~{Turok}, P.~{Boyle}, M.~{Rupen}, T.~{Landecker}, U.-L. {Pen}, and V.~{Kaspi}, \href{http://dx.doi.org/10.5281/zenodo.3765414}{``{The Canadian Hydrogen Observatory and Radio-transient Detector (CHORD)},''} in {\em Canadian Long Range Plan for Astronomy and Astrophysics White Papers}, vol.~2020, p.~28.
\newblock Oct., 2019.
\newblock \href{http://arxiv.org/abs/1911.01777}{{\ttfamily arXiv:1911.01777 [astro-ph.IM]}}.

\bibitem{HIRAX2022:Overview}
D.~{Crichton}, M.~{Aich}, A.~{Amara}, K.~{Bandura}, B.~A. {Bassett}, C.~{Bengaly}, P.~{Berner}, S.~{Bhatporia}, M.~{Bucher}, T.-C. {Chang}, H.~C. {Chiang}, J.-F. {Cliche}, C.~{Crichton}, R.~{Dave}, D.~I.~L. {De Villiers}, M.~{Dobbs}, A.~M. {Ewall-Wice}, S.~{Eyono}, C.~{Finlay}, S.~{Gaddam}, K.~{Ganga}, K.~G. {Gayley}, K.~{Gerodias}, T.~B. {Gibbon}, A.~{Gumba}, N.~{Gupta}, M.~{Harris}, H.~{Heilgendorff}, M.~{Hilton}, A.~D. {Hincks}, P.~{Hitz}, M.~{Jalilvand}, R.~P.~M. {Julie}, Z.~{Kader}, J.~{Kania}, D.~{Karagiannis}, A.~{Karastergiou}, K.~{Kesebonye}, P.~{Kittiwisit}, J.-P. {Kneib}, K.~{Knowles}, E.~R. {Kuhn}, M.~{Kunz}, R.~{Maartens}, V.~{MacKay}, S.~{MacPherson}, C.~{Monstein}, K.~{Moodley}, V.~{Mugundhan}, W.~{Naidoo}, A.~{Naidu}, L.~B. {Newburgh}, V.~{Nistane}, A.~{Di Nitto}, D.~{{\"O}l{{c}}ek}, X.~{Pan}, S.~{Paul}, J.~B. {Peterson}, E.~{Pieters}, C.~{Pieterse}, A.~{Pillay}, A.~R. {Polish}, L.~{Randrianjanahary}, A.~{Refregier}, A.~{Renard}, E.~{Retana-Montenegro}, I.~H. {Rout}, C.~{Russeeawon}, A.~V.
  {Sadr}, B.~R.~B. {Saliwanchik}, A.~{Sampath}, P.~{Sanghavi}, M.~G. {Santos}, O.~{Sengate}, J.~R. {Shaw}, J.~L. {Sievers}, O.~M. {Smirnov}, K.~M. {Smith}, U.~A.~M. {Sob}, R.~{Srianand}, P.~{Stronkhorst}, D.~D. {Sunder}, S.~{Tartakovsky}, R.~{Taylor}, P.~{Timbie}, E.~E. {Tolley}, J.~{Townsend}, W.~{Tyndall}, C.~{Ungerer}, J.~{van Dyk}, G.~{van Vuuren}, K.~{Vanderlinde}, T.~{Viant}, A.~{Walters}, J.~{Wang}, A.~{Weltman}, P.~{Woudt}, D.~{Wulf}, A.~{Zavyalov}, and Z.~{Zhang}, ``{Hydrogen Intensity and Real-Time Analysis Experiment: 256-element array status and overview},'' \href{http://dx.doi.org/10.1117/1.JATIS.8.1.011019}{{\em Journal of Astronomical Telescopes, Instruments, and Systems} {\bfseries 8} (Jan., 2022) 011019}, \href{http://arxiv.org/abs/2109.13755}{{\ttfamily arXiv:2109.13755 [astro-ph.IM]}}.

\bibitem{SKA2015:HI_IM}
M.~{Santos}, P.~{Bull}, D.~{Alonso}, S.~{Camera}, P.~{Ferreira}, G.~{Bernardi}, R.~{Maartens}, M.~{Viel}, F.~{Villaescusa-Navarro}, F.~B. {Abdalla}, M.~{Jarvis}, R.~B. {Metcalf}, A.~{Pourtsidou}, and L.~{Wolz}, \href{http://dx.doi.org/10.22323/1.215.0019}{``{Cosmology from a SKA HI intensity mapping survey},''} in {\em Advancing Astrophysics with the Square Kilometre Array (AASKA14)}, p.~19.
\newblock Apr., 2015.
\newblock \href{http://arxiv.org/abs/1501.03989}{{\ttfamily arXiv:1501.03989 [astro-ph.CO]}}.

\bibitem{PAPER2010:Overview}
A.~R. {Parsons}, D.~C. {Backer}, G.~S. {Foster}, M.~C.~H. {Wright}, R.~F. {Bradley}, N.~E. {Gugliucci}, C.~R. {Parashare}, E.~E. {Benoit}, J.~E. {Aguirre}, D.~C. {Jacobs}, C.~L. {Carilli}, D.~{Herne}, M.~J. {Lynch}, J.~R. {Manley}, and D.~J. {Werthimer}, ``{The Precision Array for Probing the Epoch of Re-ionization: Eight Station Results},'' \href{http://dx.doi.org/10.1088/0004-6256/139/4/1468}{{\em \aj} {\bfseries 139} no.~4, (Apr., 2010) 1468--1480}, \href{http://arxiv.org/abs/0904.2334}{{\ttfamily arXiv:0904.2334 [astro-ph.CO]}}.

\bibitem{GMRT2013:21cmLimit}
G.~{Paciga}, J.~G. {Albert}, K.~{Bandura}, T.-C. {Chang}, Y.~{Gupta}, C.~{Hirata}, J.~{Odegova}, U.-L. {Pen}, J.~B. {Peterson}, J.~{Roy}, J.~R. {Shaw}, K.~{Sigurdson}, and T.~{Voytek}, ``{A simulation-calibrated limit on the H I power spectrum from the GMRT Epoch of Reionization experiment},'' \href{http://dx.doi.org/10.1093/mnras/stt753}{{\em \mnras} {\bfseries 433} no.~1, (July, 2013) 639--647}, \href{http://arxiv.org/abs/1301.5906}{{\ttfamily arXiv:1301.5906 [astro-ph.CO]}}.

\bibitem{GMRT2017:uGMRT_Overview}
Y.~{Gupta}, B.~{Ajithkumar}, H.~S. {Kale}, S.~{Nayak}, S.~{Sabhapathy}, S.~{Sureshkumar}, R.~V. {Swami}, J.~N. {Chengalur}, S.~K. {Ghosh}, C.~H. {Ishwara-Chandra}, B.~C. {Joshi}, N.~{Kanekar}, D.~V. {Lal}, and S.~{Roy}, ``{The upgraded GMRT: opening new windows on the radio Universe},'' \href{http://dx.doi.org/10.18520/cs/v113/i04/707-714}{{\em Current Science} {\bfseries 113} no.~4, (Aug., 2017) 707--714}.

\bibitem{MWA2013:PhaseI_Overview}
S.~J. {Tingay}, R.~{Goeke}, J.~D. {Bowman}, D.~{Emrich}, S.~M. {Ord}, D.~A. {Mitchell}, M.~F. {Morales}, T.~{Booler}, B.~{Crosse}, R.~B. {Wayth}, C.~J. {Lonsdale}, S.~{Tremblay}, D.~{Pallot}, T.~{Colegate}, A.~{Wicenec}, N.~{Kudryavtseva}, W.~{Arcus}, D.~{Barnes}, G.~{Bernardi}, F.~{Briggs}, S.~{Burns}, J.~D. {Bunton}, R.~J. {Cappallo}, B.~E. {Corey}, A.~{Deshpande}, L.~{Desouza}, B.~M. {Gaensler}, L.~J. {Greenhill}, P.~J. {Hall}, B.~J. {Hazelton}, D.~{Herne}, J.~N. {Hewitt}, M.~{Johnston-Hollitt}, D.~L. {Kaplan}, J.~C. {Kasper}, B.~B. {Kincaid}, R.~{Koenig}, E.~{Kratzenberg}, M.~J. {Lynch}, B.~{Mckinley}, S.~R. {Mcwhirter}, E.~{Morgan}, D.~{Oberoi}, J.~{Pathikulangara}, T.~{Prabu}, R.~A. {Remillard}, A.~E.~E. {Rogers}, A.~{Roshi}, J.~E. {Salah}, R.~J. {Sault}, N.~{Udaya-Shankar}, F.~{Schlagenhaufer}, K.~S. {Srivani}, J.~{Stevens}, R.~{Subrahmanyan}, M.~{Waterson}, R.~L. {Webster}, A.~R. {Whitney}, A.~{Williams}, C.~L. {Williams}, and J.~S.~B. {Wyithe}, ``{The Murchison Widefield Array: The Square Kilometre
  Array Precursor at Low Radio Frequencies},'' \href{http://dx.doi.org/10.1017/pasa.2012.007}{{\em \pasa} {\bfseries 30} (Jan., 2013) e007}, \href{http://arxiv.org/abs/1206.6945}{{\ttfamily arXiv:1206.6945 [astro-ph.IM]}}.

\bibitem{MWA2018:PhaseII_Overview}
R.~B. {Wayth}, S.~J. {Tingay}, C.~M. {Trott}, D.~{Emrich}, M.~{Johnston-Hollitt}, B.~{McKinley}, B.~M. {Gaensler}, A.~P. {Beardsley}, T.~{Booler}, B.~{Crosse}, T.~M.~O. {Franzen}, L.~{Horsley}, D.~L. {Kaplan}, D.~{Kenney}, M.~F. {Morales}, D.~{Pallot}, G.~{Sleap}, K.~{Steele}, M.~{Walker}, A.~{Williams}, C.~{Wu}, I.~H. {Cairns}, M.~D. {Filipovic}, S.~{Johnston}, T.~{Murphy}, P.~{Quinn}, L.~{Staveley-Smith}, R.~{Webster}, and J.~S.~B. {Wyithe}, ``{The Phase II Murchison Widefield Array: Design overview},'' \href{http://dx.doi.org/10.1017/pasa.2018.37}{{\em \pasa} {\bfseries 35} (Nov., 2018) e033}, \href{http://arxiv.org/abs/1809.06466}{{\ttfamily arXiv:1809.06466 [astro-ph.IM]}}.

\bibitem{LOFAR2013:Overview}
M.~P. {van Haarlem}, M.~W. {Wise}, A.~W. {Gunst}, G.~{Heald}, J.~P. {McKean}, J.~W.~T. {Hessels}, A.~G. {de Bruyn}, R.~{Nijboer}, J.~{Swinbank}, R.~{Fallows}, M.~{Brentjens}, A.~{Nelles}, R.~{Beck}, H.~{Falcke}, R.~{Fender}, J.~{H{\"o}randel}, L.~V.~E. {Koopmans}, G.~{Mann}, G.~{Miley}, H.~{R{\"o}ttgering}, B.~W. {Stappers}, R.~A.~M.~J. {Wijers}, S.~{Zaroubi}, M.~{van den Akker}, A.~{Alexov}, J.~{Anderson}, K.~{Anderson}, A.~{van Ardenne}, M.~{Arts}, A.~{Asgekar}, I.~M. {Avruch}, F.~{Batejat}, L.~{B{\"a}hren}, M.~E. {Bell}, M.~R. {Bell}, I.~{van Bemmel}, P.~{Bennema}, M.~J. {Bentum}, G.~{Bernardi}, P.~{Best}, L.~{B{\^\i}rzan}, A.~{Bonafede}, A.~J. {Boonstra}, R.~{Braun}, J.~{Bregman}, F.~{Breitling}, R.~H. {van de Brink}, J.~{Broderick}, P.~C. {Broekema}, W.~N. {Brouw}, M.~{Br{\"u}ggen}, H.~R. {Butcher}, W.~{van Cappellen}, B.~{Ciardi}, T.~{Coenen}, J.~{Conway}, A.~{Coolen}, A.~{Corstanje}, S.~{Damstra}, O.~{Davies}, A.~T. {Deller}, R.~J. {Dettmar}, G.~{van Diepen}, K.~{Dijkstra}, P.~{Donker}, A.~{Doorduin},
  J.~{Dromer}, M.~{Drost}, A.~{van Duin}, J.~{Eisl{\"o}ffel}, J.~{van Enst}, C.~{Ferrari}, W.~{Frieswijk}, H.~{Gankema}, M.~A. {Garrett}, F.~{de Gasperin}, M.~{Gerbers}, E.~{de Geus}, J.~M. {Grie{\ss}meier}, T.~{Grit}, P.~{Gruppen}, J.~P. {Hamaker}, T.~{Hassall}, M.~{Hoeft}, H.~A. {Holties}, A.~{Horneffer}, A.~{van der Horst}, A.~{van Houwelingen}, A.~{Huijgen}, M.~{Iacobelli}, H.~{Intema}, N.~{Jackson}, V.~{Jelic}, A.~{de Jong}, E.~{Juette}, D.~{Kant}, A.~{Karastergiou}, A.~{Koers}, H.~{Kollen}, V.~I. {Kondratiev}, E.~{Kooistra}, Y.~{Koopman}, A.~{Koster}, M.~{Kuniyoshi}, M.~{Kramer}, G.~{Kuper}, P.~{Lambropoulos}, C.~{Law}, J.~{van Leeuwen}, J.~{Lemaitre}, M.~{Loose}, P.~{Maat}, G.~{Macario}, S.~{Markoff}, J.~{Masters}, R.~A. {McFadden}, D.~{McKay-Bukowski}, H.~{Meijering}, H.~{Meulman}, M.~{Mevius}, E.~{Middelberg}, R.~{Millenaar}, J.~C.~A. {Miller-Jones}, R.~N. {Mohan}, J.~D. {Mol}, J.~{Morawietz}, R.~{Morganti}, D.~D. {Mulcahy}, E.~{Mulder}, H.~{Munk}, L.~{Nieuwenhuis}, R.~{van Nieuwpoort}, J.~E.
  {Noordam}, M.~{Norden}, A.~{Noutsos}, A.~R. {Offringa}, H.~{Olofsson}, A.~{Omar}, E.~{Orr{\'u}}, R.~{Overeem}, H.~{Paas}, M.~{Pandey-Pommier}, V.~N. {Pandey}, R.~{Pizzo}, A.~{Polatidis}, D.~{Rafferty}, S.~{Rawlings}, W.~{Reich}, J.~P. {de Reijer}, J.~{Reitsma}, G.~A. {Renting}, P.~{Riemers}, E.~{Rol}, J.~W. {Romein}, J.~{Roosjen}, M.~{Ruiter}, A.~{Scaife}, K.~{van der Schaaf}, B.~{Scheers}, P.~{Schellart}, A.~{Schoenmakers}, G.~{Schoonderbeek}, M.~{Serylak}, A.~{Shulevski}, J.~{Sluman}, O.~{Smirnov}, C.~{Sobey}, H.~{Spreeuw}, M.~{Steinmetz}, C.~G.~M. {Sterks}, H.~J. {Stiepel}, K.~{Stuurwold}, M.~{Tagger}, Y.~{Tang}, C.~{Tasse}, I.~{Thomas}, S.~{Thoudam}, M.~C. {Toribio}, B.~{van der Tol}, O.~{Usov}, M.~{van Veelen}, A.~J. {van der Veen}, S.~{ter Veen}, J.~P.~W. {Verbiest}, R.~{Vermeulen}, N.~{Vermaas}, C.~{Vocks}, C.~{Vogt}, M.~{de Vos}, E.~{van der Wal}, R.~{van Weeren}, H.~{Weggemans}, P.~{Weltevrede}, S.~{White}, S.~J. {Wijnholds}, T.~{Wilhelmsson}, O.~{Wucknitz}, S.~{Yatawatta}, P.~{Zarka}, A.~{Zensus},
  and J.~{van Zwieten}, ``{LOFAR: The LOw-Frequency ARray},'' \href{http://dx.doi.org/10.1051/0004-6361/201220873}{{\em \aap} {\bfseries 556} (Aug., 2013) A2}, \href{http://arxiv.org/abs/1305.3550}{{\ttfamily arXiv:1305.3550 [astro-ph.IM]}}.

\bibitem{NenuFAR2012:Overview}
P.~{Zarka}, J.~N. {Girard}, M.~{Tagger}, and L.~{Denis}, ``{LSS/NenuFAR: The LOFAR Super Station project in Nan{{c}}ay},'' in {\em SF2A-2012: Proceedings of the Annual meeting of the French Society of Astronomy and Astrophysics}, S.~{Boissier}, P.~{de Laverny}, N.~{Nardetto}, R.~{Samadi}, D.~{Valls-Gabaud}, and H.~{Wozniak}, eds., pp.~687--694.
\newblock Dec., 2012.

\bibitem{HERA2017:PhaseI_Overview}
D.~R. {DeBoer}, A.~R. {Parsons}, J.~E. {Aguirre}, P.~{Alexander}, Z.~S. {Ali}, A.~P. {Beardsley}, G.~{Bernardi}, J.~D. {Bowman}, R.~F. {Bradley}, C.~L. {Carilli}, C.~{Cheng}, E.~{de Lera Acedo}, J.~S. {Dillon}, A.~{Ewall-Wice}, G.~{Fadana}, N.~{Fagnoni}, R.~{Fritz}, S.~R. {Furlanetto}, B.~{Glendenning}, B.~{Greig}, J.~{Grobbelaar}, B.~J. {Hazelton}, J.~N. {Hewitt}, J.~{Hickish}, D.~C. {Jacobs}, A.~{Julius}, M.~{Kariseb}, S.~A. {Kohn}, T.~{Lekalake}, A.~{Liu}, A.~{Loots}, D.~{MacMahon}, L.~{Malan}, C.~{Malgas}, M.~{Maree}, Z.~{Martinot}, N.~{Mathison}, E.~{Matsetela}, A.~{Mesinger}, M.~F. {Morales}, A.~R. {Neben}, N.~{Patra}, S.~{Pieterse}, J.~C. {Pober}, N.~{Razavi-Ghods}, J.~{Ringuette}, J.~{Robnett}, K.~{Rosie}, R.~{Sell}, C.~{Smith}, A.~{Syce}, M.~{Tegmark}, N.~{Thyagarajan}, P.~K.~G. {Williams}, and H.~{Zheng}, ``{Hydrogen Epoch of Reionization Array (HERA)},'' \href{http://dx.doi.org/10.1088/1538-3873/129/974/045001}{{\em \pasp} {\bfseries 129} no.~974, (Apr., 2017) 045001},
  \href{http://arxiv.org/abs/1606.07473}{{\ttfamily arXiv:1606.07473 [astro-ph.IM]}}.

\bibitem{Berkhout2024:HERA_PhaseII}
L.~M. {Berkhout}, D.~C. {Jacobs}, Z.~{Abdurashidova}, T.~{Adams}, J.~E. {Aguirre}, P.~{Alexander}, Z.~S. {Ali}, R.~{Baartman}, Y.~{Balfour}, A.~P. {Beardsley}, G.~{Bernardi}, T.~S. {Billings}, J.~D. {Bowman}, R.~F. {Bradley}, P.~{Bull}, J.~{Burba}, S.~{Carey}, C.~L. {Carilli}, K.-F. {Chen}, C.~{Cheng}, S.~{Choudhuri}, D.~R. {DeBoer}, E.~{de Lera Acedo}, M.~{Dexter}, J.~S. {Dillon}, S.~{Dynes}, N.~{Eksteen}, J.~{Ely}, A.~{Ewall-Wice}, N.~{Fagnoni}, R.~{Fritz}, S.~R. {Furlanetto}, K.~{Gale-Sides}, H.~{Garsden}, B.~K. {Gehlot}, A.~{Ghosh}, B.~{Glendenning}, A.~{Gorce}, D.~{Gorthi}, B.~{Greig}, J.~{Grobbelaar}, Z.~{Halday}, B.~J. {Hazelton}, J.~N. {Hewitt}, J.~{Hickish}, T.~{Huang}, A.~{Josaitis}, A.~{Julius}, M.~{Kariseb}, N.~S. {Kern}, J.~{Kerrigan}, H.~{Kim}, P.~{Kittiwisit}, S.~A. {Kohn}, M.~{Kolopanis}, A.~{Lanman}, P.~{La Plante}, A.~{Liu}, A.~{Loots}, Y.-Z. {Ma}, D.~H.~E. {MacMahon}, L.~{Malan}, C.~{Malgas}, K.~{Malgas}, B.~{Marero}, Z.~E. {Martinot}, A.~{Mesinger}, M.~{Molewa}, M.~F. {Morales},
  T.~{Mosiane}, S.~G. {Murray}, A.~R. {Neben}, B.~{Nikolic}, C.~{Devi Nunhokee}, H.~{Nuwegeld}, A.~R. {Parsons}, R.~{Pascua}, N.~{Patra}, S.~{Pieterse}, Y.~{Qin}, E.~{Rath}, N.~{Razavi-Ghods}, D.~{Riley}, J.~{Robnett}, K.~{Rosie}, M.~G. {Santos}, P.~{Sims}, S.~{Singh}, D.~{Storer}, H.~{Swarts}, J.~{Tan}, N.~{Thyagarajan}, P.~{van Wyngaarden}, P.~K.~G. {Williams}, H.~{Zheng}, and Z.~{Xu}, ``{Hydrogen Epoch of Reionization Array (HERA) Phase II Deployment and Commissioning},'' \href{http://dx.doi.org/10.48550/arXiv.2401.04304}{{\em arXiv e-prints} (Jan., 2024) arXiv:2401.04304}, \href{http://arxiv.org/abs/2401.04304}{{\ttfamily arXiv:2401.04304 [astro-ph.IM]}}.

\bibitem{LWA2019:21cmLimit}
M.~W. {Eastwood}, M.~M. {Anderson}, R.~M. {Monroe}, G.~{Hallinan}, M.~{Catha}, J.~{Dowell}, H.~{Garsden}, L.~J. {Greenhill}, B.~C. {Hicks}, J.~{Kocz}, D.~C. {Price}, F.~K. {Schinzel}, H.~{Vedantham}, and Y.~{Wang}, ``{The 21 cm Power Spectrum from the Cosmic Dawn: First Results from the OVRO-LWA},'' \href{http://dx.doi.org/10.3847/1538-3881/ab2629}{{\em \aj} {\bfseries 158} no.~2, (Aug., 2019) 84}, \href{http://arxiv.org/abs/1906.08943}{{\ttfamily arXiv:1906.08943 [astro-ph.CO]}}.

\bibitem{SKA2015:EoR}
L.~{Koopmans}, J.~{Pritchard}, G.~{Mellema}, J.~{Aguirre}, K.~{Ahn}, R.~{Barkana}, I.~{van Bemmel}, G.~{Bernardi}, A.~{Bonaldi}, F.~{Briggs}, A.~G. {de Bruyn}, T.~C. {Chang}, E.~{Chapman}, X.~{Chen}, B.~{Ciardi}, P.~{Dayal}, A.~{Ferrara}, A.~{Fialkov}, F.~{Fiore}, K.~{Ichiki}, I.~T. {Illiev}, S.~{Inoue}, V.~{Jelic}, M.~{Jones}, J.~{Lazio}, U.~{Maio}, S.~{Majumdar}, K.~J. {Mack}, A.~{Mesinger}, M.~F. {Morales}, A.~{Parsons}, U.~L. {Pen}, M.~{Santos}, R.~{Schneider}, B.~{Semelin}, R.~S. {de Souza}, R.~{Subrahmanyan}, T.~{Takeuchi}, H.~{Vedantham}, J.~{Wagg}, R.~{Webster}, S.~{Wyithe}, K.~K. {Datta}, and C.~{Trott}, \href{http://dx.doi.org/10.22323/1.215.0001}{``{The Cosmic Dawn and Epoch of Reionisation with SKA},''} in {\em Advancing Astrophysics with the Square Kilometre Array (AASKA14)}, p.~1.
\newblock Apr., 2015.
\newblock \href{http://arxiv.org/abs/1505.07568}{{\ttfamily arXiv:1505.07568 [astro-ph.CO]}}.

\bibitem{Chang2008:BAO_IM}
T.-C. {Chang}, U.-L. {Pen}, J.~B. {Peterson}, and P.~{McDonald}, ``{Baryon Acoustic Oscillation Intensity Mapping of Dark Energy},'' \href{http://dx.doi.org/10.1103/PhysRevLett.100.091303}{{\em \prl} {\bfseries 100} no.~9, (Mar., 2008) 091303}, \href{http://arxiv.org/abs/0709.3672}{{\ttfamily arXiv:0709.3672 [astro-ph]}}.

\bibitem{Morales2010:Review}
M.~F. {Morales} and J.~S.~B. {Wyithe}, ``{Reionization and Cosmology with 21-cm Fluctuations},'' \href{http://dx.doi.org/10.1146/annurev-astro-081309-130936}{{\em \araa} {\bfseries 48} (Sept., 2010) 127--171}, \href{http://arxiv.org/abs/0910.3010}{{\ttfamily arXiv:0910.3010 [astro-ph.CO]}}.

\bibitem{Bull2015:21cmIM}
P.~{Bull}, P.~G. {Ferreira}, P.~{Patel}, and M.~G. {Santos}, ``{Late-time Cosmology with 21 cm Intensity Mapping Experiments},'' \href{http://dx.doi.org/10.1088/0004-637X/803/1/21}{{\em \apj} {\bfseries 803} no.~1, (Apr., 2015) 21}, \href{http://arxiv.org/abs/1405.1452}{{\ttfamily arXiv:1405.1452 [astro-ph.CO]}}.

\bibitem{Mesinger2016:Review}
A.~Mesinger, ``{Understanding the Epoch of Cosmic Reionization},'' \href{http://dx.doi.org/10.1007/978-3-319-21957-8}{{\em Underst. Epoch Cosm. Reionization Challenges Prog.} {\bfseries 423} (2016) }.

\bibitem{de_Oliveira-Costa2008:GSM}
A.~{de Oliveira-Costa}, M.~{Tegmark}, B.~M. {Gaensler}, J.~{Jonas}, T.~L. {Landecker}, and P.~{Reich}, ``{A model of diffuse Galactic radio emission from 10 MHz to 100 GHz},'' \href{http://dx.doi.org/10.1111/j.1365-2966.2008.13376.x}{{\em \mnras} {\bfseries 388} no.~1, (July, 2008) 247--260}, \href{http://arxiv.org/abs/0802.1525}{{\ttfamily arXiv:0802.1525 [astro-ph]}}.

\bibitem{Zheng2017:GSM}
H.~{Zheng}, M.~{Tegmark}, J.~S. {Dillon}, D.~A. {Kim}, A.~{Liu}, A.~R. {Neben}, J.~{Jonas}, P.~{Reich}, and W.~{Reich}, ``{An improved model of diffuse galactic radio emission from 10 MHz to 5 THz},'' \href{http://dx.doi.org/10.1093/mnras/stw2525}{{\em \mnras} {\bfseries 464} no.~3, (Jan., 2017) 3486--3497}, \href{http://arxiv.org/abs/1605.04920}{{\ttfamily arXiv:1605.04920 [astro-ph.CO]}}.

\bibitem{Hurley-Walker2017:GLEAMI}
N.~{Hurley-Walker}, J.~R. {Callingham}, P.~J. {Hancock}, T.~M.~O. {Franzen}, L.~{Hindson}, A.~D. {Kapi{\'n}ska}, J.~{Morgan}, A.~R. {Offringa}, R.~B. {Wayth}, C.~{Wu}, Q.~{Zheng}, T.~{Murphy}, M.~E. {Bell}, K.~S. {Dwarakanath}, B.~{For}, B.~M. {Gaensler}, M.~{Johnston-Hollitt}, E.~{Lenc}, P.~{Procopio}, L.~{Staveley-Smith}, R.~{Ekers}, J.~D. {Bowman}, F.~{Briggs}, R.~J. {Cappallo}, A.~A. {Deshpande}, L.~{Greenhill}, B.~J. {Hazelton}, D.~L. {Kaplan}, C.~J. {Lonsdale}, S.~R. {McWhirter}, D.~A. {Mitchell}, M.~F. {Morales}, E.~{Morgan}, D.~{Oberoi}, S.~M. {Ord}, T.~{Prabu}, N.~U. {Shankar}, K.~S. {Srivani}, R.~{Subrahmanyan}, S.~J. {Tingay}, R.~L. {Webster}, A.~{Williams}, and C.~L. {Williams}, ``{GaLactic and Extragalactic All-sky Murchison Widefield Array (GLEAM) survey - I. A low-frequency extragalactic catalogue},'' \href{http://dx.doi.org/10.1093/mnras/stw2337}{{\em \mnras} {\bfseries 464} no.~1, (Jan., 2017) 1146--1167}, \href{http://arxiv.org/abs/1610.08318}{{\ttfamily arXiv:1610.08318 [astro-ph.GA]}}.

\bibitem{Hurley-Walker2019:GLEAMII}
N.~{Hurley-Walker}, P.~J. {Hancock}, T.~M.~O. {Franzen}, J.~R. {Callingham}, A.~R. {Offringa}, L.~{Hindson}, C.~{Wu}, M.~E. {Bell}, B.~Q. {For}, B.~M. {Gaensler}, M.~{Johnston-Hollitt}, A.~D. {Kapi{\'n}ska}, J.~{Morgan}, T.~{Murphy}, B.~{McKinley}, P.~{Procopio}, L.~{Staveley-Smith}, R.~B. {Wayth}, and Q.~{Zheng}, ``{GaLactic and Extragalactic All-sky Murchison Widefield Array (GLEAM) survey II: Galactic plane 345{\textdegree} < l < 67{\textdegree}, 180{\textdegree} < l < 240{\textdegree}},'' \href{http://dx.doi.org/10.1017/pasa.2019.37}{{\em \pasa} {\bfseries 36} (Nov., 2019) e047}, \href{http://arxiv.org/abs/1911.08127}{{\ttfamily arXiv:1911.08127 [astro-ph.GA]}}.

\bibitem{Datta2010:FGwedge}
A.~{Datta}, J.~D. {Bowman}, and C.~L. {Carilli}, ``{Bright Source Subtraction Requirements for Redshifted 21 cm Measurements},'' \href{http://dx.doi.org/10.1088/0004-637X/724/1/526}{{\em \apj} {\bfseries 724} no.~1, (Nov., 2010) 526--538}, \href{http://arxiv.org/abs/1005.4071}{{\ttfamily arXiv:1005.4071 [astro-ph.CO]}}.

\bibitem{Parsons2012:delay_spectrum_wedge}
A.~R. {Parsons}, J.~C. {Pober}, J.~E. {Aguirre}, C.~L. {Carilli}, D.~C. {Jacobs}, and D.~F. {Moore}, ``{A Per-baseline, Delay-spectrum Technique for Accessing the 21 cm Cosmic Reionization Signature},'' \href{http://dx.doi.org/10.1088/0004-637X/756/2/165}{{\em \apj} {\bfseries 756} no.~2, (Sept., 2012) 165}, \href{http://arxiv.org/abs/1204.4749}{{\ttfamily arXiv:1204.4749 [astro-ph.IM]}}.

\bibitem{Vedantham2012:image_wedge}
H.~{Vedantham}, N.~{Udaya Shankar}, and R.~{Subrahmanyan}, ``{Imaging the Epoch of Reionization: Limitations from Foreground Confusion and Imaging Algorithms},'' \href{http://dx.doi.org/10.1088/0004-637X/745/2/176}{{\em \apj} {\bfseries 745} no.~2, (Feb., 2012) 176}, \href{http://arxiv.org/abs/1106.1297}{{\ttfamily arXiv:1106.1297 [astro-ph.IM]}}.

\bibitem{Trott2012:wedge}
C.~M. {Trott}, R.~B. {Wayth}, and S.~J. {Tingay}, ``{The Impact of Point-source Subtraction Residuals on 21 cm Epoch of Reionization Estimation},'' \href{http://dx.doi.org/10.1088/0004-637X/757/1/101}{{\em \apj} {\bfseries 757} no.~1, (Sept., 2012) 101}, \href{http://arxiv.org/abs/1208.0646}{{\ttfamily arXiv:1208.0646 [astro-ph.CO]}}.

\bibitem{Morales2012:wedge}
M.~F. {Morales}, B.~{Hazelton}, I.~{Sullivan}, and A.~{Beardsley}, ``{Four Fundamental Foreground Power Spectrum Shapes for 21 cm Cosmology Observations},'' \href{http://dx.doi.org/10.1088/0004-637X/752/2/137}{{\em \apj} {\bfseries 752} no.~2, (June, 2012) 137}, \href{http://arxiv.org/abs/1202.3830}{{\ttfamily arXiv:1202.3830 [astro-ph.IM]}}.

\bibitem{Hazelton2013:wedge}
B.~J. {Hazelton}, M.~F. {Morales}, and I.~S. {Sullivan}, ``{The Fundamental Multi-baseline Mode-mixing Foreground in 21 cm Epoch of Reionization Observations},'' \href{http://dx.doi.org/10.1088/0004-637X/770/2/156}{{\em \apj} {\bfseries 770} no.~2, (June, 2013) 156}, \href{http://arxiv.org/abs/1301.3126}{{\ttfamily arXiv:1301.3126 [astro-ph.IM]}}.

\bibitem{Thyagarajan2013:wedge}
N.~{Thyagarajan}, N.~{Udaya Shankar}, R.~{Subrahmanyan}, W.~{Arcus}, G.~{Bernardi}, J.~D. {Bowman}, F.~{Briggs}, J.~D. {Bunton}, R.~J. {Cappallo}, B.~E. {Corey}, L.~{deSouza}, D.~{Emrich}, B.~M. {Gaensler}, R.~F. {Goeke}, L.~J. {Greenhill}, B.~J. {Hazelton}, D.~{Herne}, J.~N. {Hewitt}, M.~{Johnston-Hollitt}, D.~L. {Kaplan}, J.~C. {Kasper}, B.~B. {Kincaid}, R.~{Koenig}, E.~{Kratzenberg}, C.~J. {Lonsdale}, M.~J. {Lynch}, S.~R. {McWhirter}, D.~A. {Mitchell}, M.~F. {Morales}, E.~H. {Morgan}, D.~{Oberoi}, S.~M. {Ord}, J.~{Pathikulangara}, R.~A. {Remillard}, A.~E.~E. {Rogers}, D.~{Anish Roshi}, J.~E. {Salah}, R.~J. {Sault}, K.~S. {Srivani}, J.~B. {Stevens}, P.~{Thiagaraj}, S.~J. {Tingay}, R.~B. {Wayth}, M.~{Waterson}, R.~L. {Webster}, A.~R. {Whitney}, A.~J. {Williams}, C.~L. {Williams}, and J.~S.~B. {Wyithe}, ``{A Study of Fundamental Limitations to Statistical Detection of Redshifted H I from the Epoch of Reionization},'' \href{http://dx.doi.org/10.1088/0004-637X/776/1/6}{{\em \apj} {\bfseries 776} no.~1, (Oct.,
  2013) 6}, \href{http://arxiv.org/abs/1308.0565}{{\ttfamily arXiv:1308.0565 [astro-ph.CO]}}.

\bibitem{Liu2014:EoR_WindowI}
A.~{Liu}, A.~R. {Parsons}, and C.~M. {Trott}, ``{Epoch of reionization window. I. Mathematical formalism},'' \href{http://dx.doi.org/10.1103/PhysRevD.90.023018}{{\em \prd} {\bfseries 90} no.~2, (July, 2014) 023018}, \href{http://arxiv.org/abs/1404.2596}{{\ttfamily arXiv:1404.2596 [astro-ph.CO]}}.

\bibitem{Liu2011:QE}
A.~{Liu} and M.~{Tegmark}, ``{A method for 21 cm power spectrum estimation in the presence of foregrounds},'' \href{http://dx.doi.org/10.1103/PhysRevD.83.103006}{{\em \prd} {\bfseries 83} no.~10, (May, 2011) 103006}, \href{http://arxiv.org/abs/1103.0281}{{\ttfamily arXiv:1103.0281 [astro-ph.CO]}}.

\bibitem{Chapman2012:FastICA}
E.~{Chapman}, F.~B. {Abdalla}, G.~{Harker}, V.~{Jeli{\'c}}, P.~{Labropoulos}, S.~{Zaroubi}, M.~A. {Brentjens}, A.~G. {de Bruyn}, and L.~V.~E. {Koopmans}, ``{Foreground removal using FASTICA: a showcase of LOFAR-EoR},'' \href{http://dx.doi.org/10.1111/j.1365-2966.2012.21065.x}{{\em \mnras} {\bfseries 423} no.~3, (July, 2012) 2518--2532}, \href{http://arxiv.org/abs/1201.2190}{{\ttfamily arXiv:1201.2190 [astro-ph.CO]}}.

\bibitem{Ghosh2015:Bayesian_21cm}
A.~{Ghosh}, L.~V.~E. {Koopmans}, E.~{Chapman}, and V.~{Jeli{\'c}}, ``{A Bayesian analysis of redshifted 21-cm H I signal and foregrounds: simulations for LOFAR},'' \href{http://dx.doi.org/10.1093/mnras/stv1355}{{\em \mnras} {\bfseries 452} no.~2, (Sept., 2015) 1587--1600}, \href{http://arxiv.org/abs/1506.04982}{{\ttfamily arXiv:1506.04982 [astro-ph.CO]}}.

\bibitem{Ewall-Wice2021:DAYNENU}
A.~{Ewall-Wice}, N.~{Kern}, J.~S. {Dillon}, A.~{Liu}, A.~{Parsons}, S.~{Singh}, A.~{Lanman}, P.~{La Plante}, N.~{Fagnoni}, E.~d.~L. {Acedo}, D.~R. {DeBoer}, C.~{Nunhokee}, P.~{Bull}, T.-C. {Chang}, T.~J.~W. {Lazio}, J.~{Aguirre}, and S.~{Weinberg}, ``{DAYENU: a simple filter of smooth foregrounds for intensity mapping power spectra},'' \href{http://dx.doi.org/10.1093/mnras/staa3293}{{\em \mnras} {\bfseries 500} no.~4, (Jan., 2021) 5195--5213}, \href{http://arxiv.org/abs/2004.11397}{{\ttfamily arXiv:2004.11397 [astro-ph.CO]}}.

\bibitem{Mertens2018:GPR}
F.~G. {Mertens}, A.~{Ghosh}, and L.~V.~E. {Koopmans}, ``{Statistical 21-cm signal separation via Gaussian Process Regression analysis},'' \href{http://dx.doi.org/10.1093/mnras/sty1207}{{\em \mnras} {\bfseries 478} no.~3, (Aug., 2018) 3640--3652}, \href{http://arxiv.org/abs/1711.10834}{{\ttfamily arXiv:1711.10834 [astro-ph.CO]}}.

\bibitem{Kern2021:GPR}
N.~S. {Kern} and A.~{Liu}, ``{Gaussian process foreground subtraction and power spectrum estimation for 21 cm cosmology},'' \href{http://dx.doi.org/10.1093/mnras/staa3736}{{\em \mnras} {\bfseries 501} no.~1, (Feb., 2021) 1463--1480}, \href{http://arxiv.org/abs/2010.15892}{{\ttfamily arXiv:2010.15892 [astro-ph.CO]}}.

\bibitem{Chege2022:MWA_foreground}
J.~K. {Chege}, C.~H. {Jordan}, C.~{Lynch}, C.~M. {Trott}, J.~L.~B. {Line}, B.~{Pindor}, and S.~{Yoshiura}, ``{Optimising MWA EoR data processing for improved 21-cm power spectrum measurements{\textemdash}fine-tuning ionospheric corrections},'' \href{http://dx.doi.org/10.1017/pasa.2022.34}{{\em \pasa} {\bfseries 39} (Oct., 2022) e047}, \href{http://arxiv.org/abs/2207.12090}{{\ttfamily arXiv:2207.12090 [astro-ph.CO]}}.

\bibitem{Wang2024:Foreground_subtraction}
H.~{Wang}, K.~{Masui}, K.~{Bandura}, A.~{Chakraborty}, M.~{Dobbs}, S.~{Foreman}, L.~{Gray}, M.~{Halpern}, A.~{Joseph}, J.~{MacEachern}, J.~{Mena-Parra}, K.~{Miller}, L.~{Newburgh}, S.~{Paul}, A.~{Reda}, P.~{Sanghavi}, S.~{Siegel}, and D.~{Wulf}, ``{Demonstration of hybrid foreground removal on CHIME data},'' \href{http://dx.doi.org/10.48550/arXiv.2408.08949}{{\em arXiv e-prints} (Aug., 2024) arXiv:2408.08949}, \href{http://arxiv.org/abs/2408.08949}{{\ttfamily arXiv:2408.08949 [astro-ph.CO]}}.

\bibitem{Mertens2024:GPR}
F.~G. {Mertens}, J.~{Bobin}, and I.~P. {Carucci}, ``{Retrieving the 21-cm signal from the Epoch of Reionization with learnt Gaussian process kernels},'' \href{http://dx.doi.org/10.1093/mnras/stad3430}{{\em \mnras} {\bfseries 527} no.~2, (Jan., 2024) 3517--3531}, \href{http://arxiv.org/abs/2307.13545}{{\ttfamily arXiv:2307.13545 [astro-ph.CO]}}.

\bibitem{Barry2016:CalibrationError}
N.~{Barry}, B.~{Hazelton}, I.~{Sullivan}, M.~F. {Morales}, and J.~C. {Pober}, ``{Calibration requirements for detecting the 21 cm epoch of reionization power spectrum and implications for the SKA},'' \href{http://dx.doi.org/10.1093/mnras/stw1380}{{\em \mnras} {\bfseries 461} no.~3, (Sept., 2016) 3135--3144}, \href{http://arxiv.org/abs/1603.00607}{{\ttfamily arXiv:1603.00607 [astro-ph.IM]}}.

\bibitem{Patil2016:CalibrationError}
A.~H. {Patil}, S.~{Yatawatta}, S.~{Zaroubi}, L.~V.~E. {Koopmans}, A.~G. {de Bruyn}, V.~{Jeli{\'c}}, B.~{Ciardi}, I.~T. {Iliev}, M.~{Mevius}, V.~N. {Pandey}, and B.~K. {Gehlot}, ``{Systematic biases in low-frequency radio interferometric data due to calibration: the LOFAR-EoR case},'' \href{http://dx.doi.org/10.1093/mnras/stw2277}{{\em \mnras} {\bfseries 463} no.~4, (Dec., 2016) 4317--4330}, \href{http://arxiv.org/abs/1605.07619}{{\ttfamily arXiv:1605.07619 [astro-ph.IM]}}.

\bibitem{Ewall-Wice2017:CalibrationError}
A.~{Ewall-Wice}, J.~S. {Dillon}, A.~{Liu}, and J.~{Hewitt}, ``{The impact of modelling errors on interferometer calibration for 21 cm power spectra},'' \href{http://dx.doi.org/10.1093/mnras/stx1221}{{\em \mnras} {\bfseries 470} no.~2, (Sept., 2017) 1849--1870}, \href{http://arxiv.org/abs/1610.02689}{{\ttfamily arXiv:1610.02689 [astro-ph.CO]}}.

\bibitem{Bryne2019:CalibrationError}
R.~{Byrne}, M.~F. {Morales}, B.~{Hazelton}, W.~{Li}, N.~{Barry}, A.~P. {Beardsley}, R.~{Joseph}, J.~{Pober}, I.~{Sullivan}, and C.~{Trott}, ``{Fundamental Limitations on the Calibration of Redundant 21 cm Cosmology Instruments and Implications for HERA and the SKA},'' \href{http://dx.doi.org/10.3847/1538-4357/ab107d}{{\em \apj} {\bfseries 875} no.~1, (Apr., 2019) 70}, \href{http://arxiv.org/abs/1811.01378}{{\ttfamily arXiv:1811.01378 [astro-ph.IM]}}.

\bibitem{Mouri_Sardarabadi2019:CalibrationError}
A.~{Mouri Sardarabadi} and L.~V.~E. {Koopmans}, ``{Quantifying suppression of the cosmological 21-cm signal due to direction-dependent gain calibration in radio interferometers},'' \href{http://dx.doi.org/10.1093/mnras/sty3444}{{\em \mnras} {\bfseries 483} no.~4, (Mar., 2019) 5480--5490}, \href{http://arxiv.org/abs/1809.03755}{{\ttfamily arXiv:1809.03755 [astro-ph.IM]}}.

\bibitem{Neben2016:BeamErrorMWA}
A.~R. {Neben}, J.~N. {Hewitt}, R.~F. {Bradley}, J.~S. {Dillon}, G.~{Bernardi}, J.~D. {Bowman}, F.~{Briggs}, R.~J. {Cappallo}, B.~E. {Corey}, A.~A. {Deshpande}, R.~{Goeke}, L.~J. {Greenhill}, B.~J. {Hazelton}, M.~{Johnston-Hollitt}, D.~L. {Kaplan}, C.~J. {Lonsdale}, S.~R. {McWhirter}, D.~A. {Mitchell}, M.~F. {Morales}, E.~{Morgan}, D.~{Oberoi}, S.~M. {Ord}, T.~{Prabu}, N.~{Udaya Shankar}, K.~S. {Srivani}, R.~{Subrahmanyan}, S.~J. {Tingay}, R.~B. {Wayth}, R.~L. {Webster}, A.~{Williams}, and C.~L. {Williams}, ``{Beam-forming Errors in Murchison Widefield Array Phased Array Antennas and their Effects on Epoch of Reionization Science},'' \href{http://dx.doi.org/10.3847/0004-637X/820/1/44}{{\em \apj} {\bfseries 820} no.~1, (Mar., 2016) 44}, \href{http://arxiv.org/abs/1602.05249}{{\ttfamily arXiv:1602.05249 [astro-ph.IM]}}.

\bibitem{Joseph2018:BeamError}
R.~C. {Joseph}, C.~M. {Trott}, and R.~B. {Wayth}, ``{The Bias and Uncertainty of Redundant and Sky-based Calibration Under Realistic Sky and Telescope Conditions},'' \href{http://dx.doi.org/10.3847/1538-3881/aaec0b}{{\em \aj} {\bfseries 156} no.~6, (Dec., 2018) 285}, \href{http://arxiv.org/abs/1810.11237}{{\ttfamily arXiv:1810.11237 [astro-ph.IM]}}.

\bibitem{Ansah-Narh2018:BeamPerturbation}
T.~{Ansah-Narh}, F.~B. {Abdalla}, O.~M. {Smirnov}, K.~M.~B. {Asad}, and J.~R. {Shaw}, ``{Simulations of systematic direction-dependent instrumental effects in intensity mapping experiments},'' \href{http://dx.doi.org/10.1093/mnras/sty2433}{{\em \mnras} {\bfseries 481} no.~2, (Dec., 2018) 2694--2710}.

\bibitem{Orosz2019:BeamVariation}
N.~{Orosz}, J.~S. {Dillon}, A.~{Ewall-Wice}, A.~R. {Parsons}, and N.~{Thyagarajan}, ``{Mitigating the effects of antenna-to-antenna variation on redundant-baseline calibration for 21 cm cosmology},'' \href{http://dx.doi.org/10.1093/mnras/stz1287}{{\em \mnras} {\bfseries 487} no.~1, (July, 2019) 537--549}, \href{http://arxiv.org/abs/1809.09728}{{\ttfamily arXiv:1809.09728 [astro-ph.CO]}}.

\bibitem{Joseph2020:BeamVariation}
R.~C. {Joseph}, C.~M. {Trott}, R.~B. {Wayth}, and A.~{Nasirudin}, ``{Calibration and 21-cm power spectrum estimation in the presence of antenna beam variations},'' \href{http://dx.doi.org/10.1093/mnras/stz3375}{{\em \mnras} {\bfseries 492} no.~2, (Feb., 2020) 2017--2028}, \href{http://arxiv.org/abs/1911.13088}{{\ttfamily arXiv:1911.13088 [astro-ph.IM]}}.

\bibitem{Choudhuri2021:BeamVariation}
S.~{Choudhuri}, P.~{Bull}, and H.~{Garsden}, ``{Patterns of primary beam non-redundancy in close-packed 21 cm array observations},'' \href{http://dx.doi.org/10.1093/mnras/stab1795}{{\em \mnras} {\bfseries 506} no.~2, (Sept., 2021) 2066--2088}, \href{http://arxiv.org/abs/2101.02684}{{\ttfamily arXiv:2101.02684 [astro-ph.CO]}}.

\bibitem{Kim2022:BeamPerturbation}
H.~{Kim}, B.~D. {Nhan}, J.~N. {Hewitt}, N.~S. {Kern}, J.~S. {Dillon}, E.~{de Lera Acedo}, S.~B.~C. {Dynes}, N.~{Mahesh}, N.~{Fagnoni}, and D.~R. {DeBoer}, ``{The Impact of Beam Variations on Power Spectrum Estimation for 21 cm Cosmology. I. Simulations of Foreground Contamination for HERA},'' \href{http://dx.doi.org/10.3847/1538-4357/ac9eaf}{{\em \apj} {\bfseries 941} no.~2, (Dec., 2022) 207}, \href{http://arxiv.org/abs/2210.16421}{{\ttfamily arXiv:2210.16421 [astro-ph.IM]}}.

\bibitem{Ewall-Wice2016:Limit_Relection}
A.~{Ewall-Wice}, J.~S. {Dillon}, J.~N. {Hewitt}, A.~{Loeb}, A.~{Mesinger}, A.~R. {Neben}, A.~R. {Offringa}, M.~{Tegmark}, N.~{Barry}, A.~P. {Beardsley}, G.~{Bernardi}, J.~D. {Bowman}, F.~{Briggs}, R.~J. {Cappallo}, P.~{Carroll}, B.~E. {Corey}, A.~{de Oliveira-Costa}, D.~{Emrich}, L.~{Feng}, B.~M. {Gaensler}, R.~{Goeke}, L.~J. {Greenhill}, B.~J. {Hazelton}, N.~{Hurley-Walker}, M.~{Johnston-Hollitt}, D.~C. {Jacobs}, D.~L. {Kaplan}, J.~C. {Kasper}, H.~{Kim}, E.~{Kratzenberg}, E.~{Lenc}, J.~{Line}, C.~J. {Lonsdale}, M.~J. {Lynch}, B.~{McKinley}, S.~R. {McWhirter}, D.~A. {Mitchell}, M.~F. {Morales}, E.~{Morgan}, N.~{Thyagarajan}, D.~{Oberoi}, S.~M. {Ord}, S.~{Paul}, B.~{Pindor}, J.~C. {Pober}, T.~{Prabu}, P.~{Procopio}, J.~{Riding}, A.~E.~E. {Rogers}, A.~{Roshi}, N.~U. {Shankar}, S.~K. {Sethi}, K.~S. {Srivani}, R.~{Subrahmanyan}, I.~S. {Sullivan}, S.~J. {Tingay}, C.~M. {Trott}, M.~{Waterson}, R.~B. {Wayth}, R.~L. {Webster}, A.~R. {Whitney}, A.~{Williams}, C.~L. {Williams}, C.~{Wu}, and J.~S.~B. {Wyithe}, ``{First
  limits on the 21 cm power spectrum during the Epoch of X-ray heating},'' \href{http://dx.doi.org/10.1093/mnras/stw1022}{{\em \mnras} {\bfseries 460} no.~4, (Aug., 2016) 4320--4347}, \href{http://arxiv.org/abs/1605.00016}{{\ttfamily arXiv:1605.00016 [astro-ph.CO]}}.

\bibitem{Kern2019:Relection_Model}
N.~S. {Kern}, A.~R. {Parsons}, J.~S. {Dillon}, A.~E. {Lanman}, N.~{Fagnoni}, and E.~{de Lera Acedo}, ``{Mitigating Internal Instrument Coupling for 21 cm Cosmology I: Temporal and Spectral Modeling in Simulations},'' \href{http://dx.doi.org/10.48550/arXiv.1909.11732}{{\em arXiv e-prints} (Sept., 2019) arXiv:1909.11732}, \href{http://arxiv.org/abs/1909.11732}{{\ttfamily arXiv:1909.11732 [astro-ph.IM]}}.

\bibitem{Ung2020:MWA_Coupling_Sim}
D.~C.~X. {Ung}, M.~{Sokolowski}, A.~T. {Sutinjo}, and D.~B. {Davidson}, ``{Noise Temperature of Phased Array Radio Telescope: The Murchison Widefield Array and the Engineering Development Array},'' \href{http://dx.doi.org/10.1109/TAP.2020.2980334}{{\em IEEE Transactions on Antennas and Propagation} {\bfseries 68} no.~7, (July, 2020) 5395--5404}, \href{http://arxiv.org/abs/2003.05116}{{\ttfamily arXiv:2003.05116 [astro-ph.IM]}}.

\bibitem{Josaitis2022:MutualCoupling}
A.~T. {Josaitis}, A.~{Ewall-Wice}, N.~{Fagnoni}, and E.~{de Lera Acedo}, ``{Array element coupling in radio interferometry I: a semi-analytic approach},'' \href{http://dx.doi.org/10.1093/mnras/stac916}{{\em \mnras} {\bfseries 514} no.~2, (Aug., 2022) 1804--1827}, \href{http://arxiv.org/abs/2110.10879}{{\ttfamily arXiv:2110.10879 [astro-ph.IM]}}.

\bibitem{Rath_Pascua2024:Mutual_Coupling}
E.~{Rath}, R.~{Pascua}, A.~T. {Josaitis}, A.~{Ewall-Wice}, N.~{Fagnoni}, E.~{de Lera Acedo}, Z.~E. {Martinot}, Z.~{Abdurashidova}, T.~{Adams}, J.~E. {Aguirre}, R.~{Baartman}, A.~P. {Beardsley}, L.~M. {Berkhout}, G.~{Bernardi}, T.~S. {Billings}, J.~D. {Bowman}, P.~{Bull}, J.~{Burba}, R.~{Byrne}, S.~{Carey}, K.~F. {Chen}, S.~{Choudhuri}, T.~{Cox}, D.~R. {DeBoer}, M.~{Dexter}, J.~S. {Dillon}, S.~{Dynes}, N.~{Eksteen}, J.~{Ely}, R.~{Fritz}, S.~R. {Furlanetto}, K.~{Gale-Sides}, H.~{Garsden}, B.~K. {Gehlot}, A.~{Ghosh}, A.~{Gorce}, D.~{Gorthi}, Z.~{Halday}, B.~J. {Hazelton}, J.~N. {Hewitt}, J.~{Hickish}, T.~{Huang}, D.~C. {Jacobs}, N.~S. {Kern}, J.~{Kerrigan}, P.~{Kittiwisit}, M.~{Kolopanis}, A.~{Lanman}, A.~{Liu}, Y.~Z. {Ma}, D.~H.~E. {MacMahon}, L.~{Malan}, C.~{Malgas}, K.~{Malgas}, B.~{Marero}, L.~{McBride}, A.~{Mesinger}, N.~{Mohamed-Hinds}, M.~{Molewa}, M.~F. {Morales}, S.~G. {Murray}, B.~{Nikolic}, H.~{Nuwegeld}, A.~R. {Parsons}, N.~{Patra}, P.~{La Plante}, Y.~{Qin}, N.~{Razavi-Ghods}, D.~{Riley},
  J.~{Robnett}, K.~{Rosie}, M.~G. {Santos}, P.~{Sims}, S.~{Singh}, D.~{Storer}, H.~{Swarts}, J.~{Tan}, M.~J. {Wilensky}, P.~K.~G. {Williams}, P.~{van Wyngaarden}, and H.~{Zheng}, ``{Investigating Mutual Coupling in the Hydrogen Epoch of Reionization Array and Mitigating its Effects on the 21-cm Power Spectrum},'' \href{http://dx.doi.org/10.48550/arXiv.2406.08549}{{\em arXiv e-prints} (June, 2024) arXiv:2406.08549}, \href{http://arxiv.org/abs/2406.08549}{{\ttfamily arXiv:2406.08549 [astro-ph.CO]}}.

\bibitem{Chen2025:RFI}
K.-F. {Chen}, M.~J. {Wilensky}, A.~{Liu}, J.~S. {Dillon}, J.~N. {Hewitt}, T.~{Adams}, J.~E. {Aguirre}, R.~{Baartman}, A.~P. {Beardsley}, L.~M. {Berkhout}, G.~{Bernardi}, T.~S. {Billings}, J.~D. {Bowman}, P.~{Bull}, J.~{Burba}, R.~{Byrne}, S.~{Carey}, S.~{Choudhuri}, T.~{Cox}, D.~R. {DeBoer}, M.~{Dexter}, N.~{Eksteen}, J.~{Ely}, A.~{Ewall-Wice}, S.~R. {Furlanetto}, K.~{Gale-Sides}, H.~{Garsden}, B.~K. {Gehlot}, A.~{Gorce}, D.~{Gorthi}, Z.~{Halday}, B.~J. {Hazelton}, J.~{Hickish}, D.~C. {Jacobs}, A.~{Josaitis}, N.~S. {Kern}, J.~{Kerrigan}, P.~{Kittiwisit}, M.~{Kolopanis}, P.~L. {Plante}, A.~{Lanman}, Y.-Z. {Ma}, D.~H.~E. {MacMahon}, L.~{Malan}, C.~{Malgas}, K.~{Malgas}, B.~{Marero}, Z.~E. {Martinot}, L.~{McBride}, A.~{Mesinger}, N.~{Mohamed-Hinds}, M.~{Molewa}, M.~F. {Morales}, S.~G. {Murray}, H.~{Nuwegeld}, A.~R. {Parsons}, R.~{Pascua}, Y.~{Qin}, E.~{Rath}, N.~{Razavi-Ghods}, J.~{Robnett}, M.~G. {Santos}, P.~{Sims}, S.~{Singh}, D.~{Storer}, H.~{Swarts}, J.~{Tan}, P.~v. {Wyngaarden}, and H.~{Zheng}, ``{Impacts
  and Statistical Mitigation of Missing Data on the 21 cm Power Spectrum: A Case Study with the Hydrogen Epoch of Reionization Array},'' \href{http://dx.doi.org/10.3847/1538-4357/ad9b91}{{\em \apj} {\bfseries 979} no.~2, (Feb., 2025) 191}, \href{http://arxiv.org/abs/2411.10529}{{\ttfamily arXiv:2411.10529 [astro-ph.CO]}}.

\bibitem{Darwish:2020prn}
O.~Darwish, S.~Foreman, M.~M. Abidi, T.~Baldauf, B.~D. Sherwin, and P.~D. Meerburg, ``{Density reconstruction from biased tracers and its application to primordial non-Gaussianity},'' \href{http://dx.doi.org/10.1103/PhysRevD.104.123520}{{\em Phys. Rev. D} {\bfseries 104} no.~12, (2021) 123520}, \href{http://arxiv.org/abs/2007.08472}{{\ttfamily arXiv:2007.08472 [astro-ph.CO]}}.

\bibitem{Wang:2023lvt}
Z.~Wang and D.~Jeong, ``{Reconstructing the long-wavelength matter density fluctuation modes from the scalar-type clustering fossils},'' \href{http://dx.doi.org/10.1088/1475-7516/2024/07/020}{{\em JCAP} {\bfseries 07} (2024) 020}, \href{http://arxiv.org/abs/2312.17321}{{\ttfamily arXiv:2312.17321 [astro-ph.CO]}}.

\bibitem{Modi:2019hnu}
C.~Modi, M.~White, A.~Slosar, and E.~Castorina, ``{Reconstructing large-scale structure with neutral hydrogen surveys},'' \href{http://dx.doi.org/10.1088/1475-7516/2019/11/023}{{\em JCAP} {\bfseries 11} (2019) 023}, \href{http://arxiv.org/abs/1907.02330}{{\ttfamily arXiv:1907.02330 [astro-ph.CO]}}.

\bibitem{Nguyen:2024yth}
N.-M. Nguyen, F.~Schmidt, B.~Tucci, M.~Reinecke, and A.~Kosti\'c, ``{How much information can be extracted from galaxy clustering at the field level?},'' \href{http://arxiv.org/abs/2403.03220}{{\ttfamily arXiv:2403.03220 [astro-ph.CO]}}.

\bibitem{Jasche:2012kq}
J.~Jasche and B.~D. Wandelt, ``{Bayesian physical reconstruction of initial conditions from large scale structure surveys},'' \href{http://dx.doi.org/10.1093/mnras/stt449}{{\em Mon. Not. Roy. Astron. Soc.} {\bfseries 432} (2013) 894}, \href{http://arxiv.org/abs/1203.3639}{{\ttfamily arXiv:1203.3639 [astro-ph.CO]}}.

\bibitem{Kitaura:2012tu}
F.-S. Kitaura, ``{The Initial Conditions of the Universe from Constrained Simulations},'' \href{http://dx.doi.org/10.1093/mnrasl/sls029}{{\em Mon. Not. Roy. Astron. Soc.} {\bfseries 429} (2013) 84}, \href{http://arxiv.org/abs/1203.4184}{{\ttfamily arXiv:1203.4184 [astro-ph.CO]}}.

\bibitem{Wang:2014hia}
H.~Wang, H.~J. Mo, X.~Yang, Y.~P. Jing, and W.~P. Lin, ``{ELUCID - Exploring the Local Universe with reConstructed Initial Density field I: Hamiltonian Markov Chain Monte Carlo Method with Particle Mesh Dynamics},'' \href{http://dx.doi.org/10.1088/0004-637X/794/1/94}{{\em Astrophys. J.} {\bfseries 794} no.~1, (2014) 94}, \href{http://arxiv.org/abs/1407.3451}{{\ttfamily arXiv:1407.3451 [astro-ph.CO]}}.

\bibitem{Jasche:2014vpa}
J.~Jasche, F.~Leclercq, and B.~D. Wandelt, ``{Past and present cosmic structure in the SDSS DR7 main sample},'' \href{http://dx.doi.org/10.1088/1475-7516/2015/01/036}{{\em JCAP} {\bfseries 01} (2015) 036}, \href{http://arxiv.org/abs/1409.6308}{{\ttfamily arXiv:1409.6308 [astro-ph.CO]}}.

\bibitem{Ata:2014ssa}
M.~Ata, F.-S. Kitaura, and V.~M\"uller, ``{Bayesian inference of cosmic density fields from non-linear, scale-dependent, and stochastic biased tracers},'' \href{http://dx.doi.org/10.1093/mnras/stu2347}{{\em Mon. Not. Roy. Astron. Soc.} {\bfseries 446} no.~4, (2015) 4250--4259}, \href{http://arxiv.org/abs/1408.2566}{{\ttfamily arXiv:1408.2566 [astro-ph.CO]}}.

\bibitem{Wang:2016qbz}
H.~Wang, H.~J. Mo, X.~Yang, Y.~Zhang, J.~Shi, Y.~P. Jing, C.~Liu, S.~Li, X.~Kang, and Y.~Gao, ``{ELUCID - Exploring the Local Universe with reConstructed Initial Density field III: Constrained Simulation in the SDSS Volume},'' \href{http://dx.doi.org/10.3847/0004-637X/831/2/164}{{\em Astrophys. J.} {\bfseries 831} no.~2, (2016) 164}, \href{http://arxiv.org/abs/1608.01763}{{\ttfamily arXiv:1608.01763 [astro-ph.CO]}}.

\bibitem{BOSS:2016ipy}
{\bfseries BOSS} Collaboration, M.~Ata {\em et~al.}, ``{The Clustering of Galaxies in the Completed SDSS-III Baryon Oscillation Spectroscopic Survey: Cosmic Flows and Cosmic Web from Luminous Red Galaxies},'' \href{http://dx.doi.org/10.1093/mnras/stx178}{{\em Mon. Not. Roy. Astron. Soc.} {\bfseries 467} no.~4, (2017) 3993--4014}, \href{http://arxiv.org/abs/1605.09745}{{\ttfamily arXiv:1605.09745 [astro-ph.CO]}}.

\bibitem{Modi:2018cfi}
C.~Modi, Y.~Feng, and U.~Seljak, ``{Cosmological Reconstruction From Galaxy Light: Neural Network Based Light-Matter Connection},'' \href{http://dx.doi.org/10.1088/1475-7516/2018/10/028}{{\em JCAP} {\bfseries 10} (2018) 028}, \href{http://arxiv.org/abs/1805.02247}{{\ttfamily arXiv:1805.02247 [astro-ph.CO]}}.

\bibitem{Jasche:2018oym}
J.~Jasche and G.~Lavaux, ``{Physical Bayesian modelling of the non-linear matter distribution: new insights into the Nearby Universe},'' \href{http://dx.doi.org/10.1051/0004-6361/201833710}{{\em Astron. Astrophys.} {\bfseries 625} (2019) A64}, \href{http://arxiv.org/abs/1806.11117}{{\ttfamily arXiv:1806.11117 [astro-ph.CO]}}.

\bibitem{Lavaux:2019fjr}
G.~Lavaux, J.~Jasche, and F.~Leclercq, ``{Systematic-free inference of the cosmic matter density field from SDSS3-BOSS data},'' \href{http://arxiv.org/abs/1909.06396}{{\ttfamily arXiv:1909.06396 [astro-ph.CO]}}.

\bibitem{Schmidt:2018bkr}
F.~Schmidt, F.~Elsner, J.~Jasche, N.~M. Nguyen, and G.~Lavaux, ``{A rigorous EFT-based forward model for large-scale structure},'' \href{http://dx.doi.org/10.1088/1475-7516/2019/01/042}{{\em JCAP} {\bfseries 01} (2019) 042}, \href{http://arxiv.org/abs/1808.02002}{{\ttfamily arXiv:1808.02002 [astro-ph.CO]}}.

\bibitem{Cabass:2019lqx}
G.~Cabass and F.~Schmidt, ``{The EFT Likelihood for Large-Scale Structure},'' \href{http://dx.doi.org/10.1088/1475-7516/2020/04/042}{{\em JCAP} {\bfseries 04} (2020) 042}, \href{http://arxiv.org/abs/1909.04022}{{\ttfamily arXiv:1909.04022 [astro-ph.CO]}}.

\bibitem{Cabass:2020nwf}
G.~Cabass and F.~Schmidt, ``{The Likelihood for LSS: Stochasticity of Bias Coefficients at All Orders},'' \href{http://dx.doi.org/10.1088/1475-7516/2020/07/051}{{\em JCAP} {\bfseries 07} (2020) 051}, \href{http://arxiv.org/abs/2004.00617}{{\ttfamily arXiv:2004.00617 [astro-ph.CO]}}.

\bibitem{Cabass:2020jqo}
G.~Cabass, ``{The EFT Likelihood for Large-Scale Structure in Redshift Space},'' \href{http://dx.doi.org/10.1088/1475-7516/2021/01/067}{{\em JCAP} {\bfseries 01} (2021) 067}, \href{http://arxiv.org/abs/2007.14988}{{\ttfamily arXiv:2007.14988 [astro-ph.CO]}}.

\bibitem{Elsner:2019rql}
F.~Elsner, F.~Schmidt, J.~Jasche, G.~Lavaux, and N.-M. Nguyen, ``{Cosmology inference from a biased density field using the EFT-based likelihood},'' \href{http://dx.doi.org/10.1088/1475-7516/2020/01/029}{{\em JCAP} {\bfseries 01} (2020) 029}, \href{http://arxiv.org/abs/1906.07143}{{\ttfamily arXiv:1906.07143 [astro-ph.CO]}}.

\bibitem{Schmidt:2020tao}
F.~Schmidt, ``{Sigma-Eight at the Percent Level: The EFT Likelihood in Real Space},'' \href{http://dx.doi.org/10.1088/1475-7516/2021/04/032}{{\em JCAP} {\bfseries 04} (2021) 032}, \href{http://arxiv.org/abs/2009.14176}{{\ttfamily arXiv:2009.14176 [astro-ph.CO]}}.

\bibitem{Modi:2022pzm}
C.~Modi, Y.~Li, and D.~Blei, ``{Reconstructing the universe with variational self-boosted sampling},'' \href{http://dx.doi.org/10.1088/1475-7516/2023/03/059}{{\em JCAP} {\bfseries 03} (2023) 059}, \href{http://arxiv.org/abs/2206.15433}{{\ttfamily arXiv:2206.15433 [astro-ph.IM]}}.

\bibitem{Dai:2022dso}
B.~Dai and U.~Seljak, ``{Translation and rotation equivariant normalizing flow (TRENF) for optimal cosmological analysis},'' \href{http://dx.doi.org/10.1093/mnras/stac2010}{{\em Mon. Not. Roy. Astron. Soc.} {\bfseries 516} no.~2, (2022) 2363--2373}, \href{http://arxiv.org/abs/2202.05282}{{\ttfamily arXiv:2202.05282 [astro-ph.CO]}}.

\bibitem{Charnock:2019rbk}
T.~Charnock, G.~Lavaux, B.~D. Wandelt, S.~Sarma~Boruah, J.~Jasche, and M.~J. Hudson, ``{Neural physical engines for inferring the halo mass distribution function},'' \href{http://dx.doi.org/10.1093/mnras/staa682}{{\em Mon. Not. Roy. Astron. Soc.} {\bfseries 494} no.~1, (2020) 50--61}, \href{http://arxiv.org/abs/1909.06379}{{\ttfamily arXiv:1909.06379 [astro-ph.CO]}}.

\bibitem{Doeser:2023yzv}
L.~Doeser, D.~Jamieson, S.~Stopyra, G.~Lavaux, F.~Leclercq, and J.~Jasche, ``{Bayesian Inference of Initial Conditions from Non-Linear Cosmic Structures using Field-Level Emulators},'' \href{http://arxiv.org/abs/2312.09271}{{\ttfamily arXiv:2312.09271 [astro-ph.CO]}}.

\bibitem{Kostic:2022vok}
A.~Kosti\'c, N.-M. Nguyen, F.~Schmidt, and M.~Reinecke, ``{Consistency tests of field level inference with the EFT likelihood},'' \href{http://dx.doi.org/10.1088/1475-7516/2023/07/063}{{\em JCAP} {\bfseries 07} (2023) 063}, \href{http://arxiv.org/abs/2212.07875}{{\ttfamily arXiv:2212.07875 [astro-ph.CO]}}.

\bibitem{Bayer:2023rmj}
A.~E. Bayer, U.~Seljak, and C.~Modi, ``{Field-Level Inference with Microcanonical Langevin Monte Carlo},'' in {\em {40th International Conference on Machine Learning}}.
\newblock 7, 2023.
\newblock \href{http://arxiv.org/abs/2307.09504}{{\ttfamily arXiv:2307.09504 [astro-ph.CO]}}.

\bibitem{Leclercq:2021ctr}
F.~Leclercq and A.~Heavens, ``{On the accuracy and precision of correlation functions and field-level inference in cosmology},'' \href{http://dx.doi.org/10.1093/mnrasl/slab081}{{\em Mon. Not. Roy. Astron. Soc.} {\bfseries 506} no.~1, (2021) L85--L90}, \href{http://arxiv.org/abs/2103.04158}{{\ttfamily arXiv:2103.04158 [astro-ph.CO]}}.

\bibitem{Andrews:2022nvv}
A.~Andrews, J.~Jasche, G.~Lavaux, and F.~Schmidt, ``{Bayesian field-level inference of primordial non-Gaussianity using next-generation galaxy surveys},'' \href{http://dx.doi.org/10.1093/mnras/stad432}{{\em Mon. Not. Roy. Astron. Soc.} {\bfseries 520} no.~4, (2023) 5746--5763}, \href{http://arxiv.org/abs/2203.08838}{{\ttfamily arXiv:2203.08838 [astro-ph.CO]}}.

\bibitem{Porqueres:2023drp}
N.~Porqueres, A.~Heavens, D.~Mortlock, G.~Lavaux, and T.~L. Makinen, ``{Field-level inference of cosmic shear with intrinsic alignments and baryons},'' \href{http://arxiv.org/abs/2304.04785}{{\ttfamily arXiv:2304.04785 [astro-ph.CO]}}.

\bibitem{Beyond-2pt:2024mqz}
{\bfseries Beyond-2pt} Collaboration, E.~Krause {\em et~al.}, ``{A Parameter-Masked Mock Data Challenge for Beyond-Two-Point Galaxy Clustering Statistics},'' \href{http://arxiv.org/abs/2405.02252}{{\ttfamily arXiv:2405.02252 [astro-ph.CO]}}.

\bibitem{Sullivan:2024jxe}
J.~M. Sullivan and S.-F. Chen, ``{Local Primordial Non-Gaussian Bias at the Field Level},'' \href{http://arxiv.org/abs/2410.18039}{{\ttfamily arXiv:2410.18039 [astro-ph.CO]}}.

\bibitem{Taruya:2021jhg}
A.~Taruya and K.~Akitsu, ``{Lagrangian approach to super-sample effects on biased tracers at field level: galaxy density fields and intrinsic alignments},'' \href{http://dx.doi.org/10.1088/1475-7516/2021/11/061}{{\em JCAP} {\bfseries 11} no.~11, (2021) 061}, \href{http://arxiv.org/abs/2106.04789}{{\ttfamily arXiv:2106.04789 [astro-ph.CO]}}.

\bibitem{Qin:2022xho}
W.~Qin, K.~Schutz, A.~Smith, E.~Garaldi, R.~Kannan, T.~R. Slatyer, and M.~Vogelsberger, ``{Effective bias expansion for 21-cm cosmology in redshift space},'' \href{http://dx.doi.org/10.1103/PhysRevD.106.123506}{{\em Phys. Rev. D} {\bfseries 106} no.~12, (2022) 123506}, \href{http://arxiv.org/abs/2205.06270}{{\ttfamily arXiv:2205.06270 [astro-ph.CO]}}.

\bibitem{McQuinn:2018zwa}
M.~McQuinn and A.~D'Aloisio, ``{The observable 21cm signal from reionization may be perturbative},'' \href{http://dx.doi.org/10.1088/1475-7516/2018/10/016}{{\em JCAP} {\bfseries 10} (2018) 016}, \href{http://arxiv.org/abs/1806.08372}{{\ttfamily arXiv:1806.08372 [astro-ph.CO]}}.

\bibitem{Qin:2025olv}
W.~Qin, K.~Schutz, O.~Rosenstein, S.~O'Neil, and M.~Vogelsberger, ``{Supersizing hydrodynamical simulations of reionization using perturbative techniques},'' \href{http://arxiv.org/abs/2504.02929}{{\ttfamily arXiv:2504.02929 [astro-ph.CO]}}.

\bibitem{Qin:2025lyy}
W.~Qin, K.-F. Chen, K.~Schutz, and A.~Liu, ``{Effective bias expansion for circumventing 21 cm foregrounds},'' \href{http://arxiv.org/abs/2508.13268}{{\ttfamily arXiv:2508.13268 [astro-ph.CO]}}.

\bibitem{2021arXiv210700630K}
D.~P. {Kingma}, T.~{Salimans}, B.~{Poole}, and J.~{Ho}, ``{Variational Diffusion Models},'' \href{http://dx.doi.org/10.48550/arXiv.2107.00630}{{\em arXiv e-prints} (July, 2021) arXiv:2107.00630}, \href{http://arxiv.org/abs/2107.00630}{{\ttfamily arXiv:2107.00630 [cs.LG]}}.

\bibitem{Ono:2024jhn}
V.~Ono, C.~F. Park, N.~Mudur, Y.~Ni, C.~Cuesta-Lazaro, and F.~Villaescusa-Navarro, ``{Debiasing with Diffusion: Probabilistic reconstruction of Dark Matter fields from galaxies with CAMELS},'' \href{http://arxiv.org/abs/2403.10648}{{\ttfamily arXiv:2403.10648 [astro-ph.CO]}}.

\bibitem{Park:2023ync}
C.~F. Park, V.~Ono, N.~Mudur, Y.~Ni, and C.~Cuesta-Lazaro, ``{Probabilistic reconstruction of Dark Matter fields from biased tracers using diffusion models},'' \href{http://arxiv.org/abs/2311.08558}{{\ttfamily arXiv:2311.08558 [astro-ph.CO]}}.

\bibitem{park2024d}
C.~F. Park, N.~Mudur, C.~Cuesta-Lazaro, Y.~Ni, V.~Ono, and D.~Finkbeiner, ``3d reconstruction of dark matter fields with diffusion models: Towards application to galaxy surveys,'' in {\em ICML 2024 AI for Science Workshop}.
\newblock 2024.
\newblock \url{https://openreview.net/forum?id=7k2Eh7OCoz}.

\bibitem{Legin:2023jxc}
R.~Legin, M.~Ho, P.~Lemos, L.~Perreault-Levasseur, S.~Ho, Y.~Hezaveh, and B.~Wandelt, ``{Posterior sampling of the initial conditions of the universe from non-linear large scale structures using score-based generative models},'' \href{http://dx.doi.org/10.1093/mnrasl/slad152}{{\em Mon. Not. Roy. Astron. Soc.} {\bfseries 527} no.~1, (2023) L173--L178}, \href{http://arxiv.org/abs/2304.03788}{{\ttfamily arXiv:2304.03788 [astro-ph.CO]}}.

\bibitem{Kennedy:2023zos}
J.~Kennedy, J.~C. Carr, S.~Gagnon-Hartman, A.~Liu, J.~Mirocha, and Y.~Cui, ``{Machine-learning recovery of foreground wedge-removed 21-cm light cones for high-z galaxy mapping},'' \href{http://dx.doi.org/10.1093/mnras/stae760}{{\em Mon. Not. Roy. Astron. Soc.} {\bfseries 529} no.~4, (2024) 3684--3698}, \href{http://arxiv.org/abs/2308.09740}{{\ttfamily arXiv:2308.09740 [astro-ph.CO]}}.

\bibitem{Li:2019znt}
W.~Li, H.~Xu, Z.~Ma, R.~Zhu, D.~Hu, Z.~Zhu, J.~Gu, C.~Shan, J.~Zhu, and X.-P. Wu, ``{Separating the EoR Signal with a Convolutional Denoising Autoencoder: A Deep-learning-based Method},'' \href{http://dx.doi.org/10.1093/mnras/stz582}{{\em Mon. Not. Roy. Astron. Soc.} {\bfseries 485} no.~2, (2019) 2628--2637}, \href{http://arxiv.org/abs/1902.09278}{{\ttfamily arXiv:1902.09278 [astro-ph.IM]}}.

\bibitem{Makinen:2020gvh}
T.~L. Makinen, L.~Lancaster, F.~Villaescusa-Navarro, P.~Melchior, S.~Ho, L.~Perreault-Levasseur, and D.~N. Spergel, ``{deep21: a deep learning method for 21 cm foreground removal},'' \href{http://dx.doi.org/10.1088/1475-7516/2021/04/081}{{\em JCAP} {\bfseries 04} (2021) 081}, \href{http://arxiv.org/abs/2010.15843}{{\ttfamily arXiv:2010.15843 [astro-ph.CO]}}.

\bibitem{Gagnon-Hartman:2021erd}
S.~Gagnon-Hartman, Y.~Cui, A.~Liu, S.~Ravanbakhsh, and J.~Kennedy, ``{Recovering the Wedge Modes Lost to 21-cm Foregrounds},'' \href{http://dx.doi.org/10.1093/mnras/stab1158}{{\em Mon. Not. Roy. Astron. Soc.} {\bfseries 504} no.~4, (2021) 4716--4729}, \href{http://arxiv.org/abs/2102.08382}{{\ttfamily arXiv:2102.08382 [astro-ph.CO]}}. [Erratum: Mon.Not.Roy.Astron.Soc. 529, 2539--2542 (2024)].

\bibitem{Bianco:2023eec}
M.~Bianco, S.~K. Giri, D.~Prelogovi\'c, T.~Chen, F.~G. Mertens, E.~Tolley, A.~Mesinger, and J.-P. Kneib, ``{Deep learning approach for identification of H~ii regions during reionization in 21-cm observations \textendash{} II. Foreground contamination},'' \href{http://dx.doi.org/10.1093/mnras/stae257}{{\em Mon. Not. Roy. Astron. Soc.} {\bfseries 528} no.~3, (2024) 5212--5230}, \href{http://arxiv.org/abs/2304.02661}{{\ttfamily arXiv:2304.02661 [astro-ph.IM]}}.

\bibitem{Bianco:2024jhe}
M.~Bianco, S.~K. Giri, R.~Sharma, T.~Chen, S.~P. Krishna, C.~Finlay, V.~Nistane, P.~Denzel, M.~De~Santis, and H.~Ghorbel, ``{Deep learning approach for identification of HII regions during reionization in 21-cm observations -- III. image recovery},'' \href{http://arxiv.org/abs/2408.16814}{{\ttfamily arXiv:2408.16814 [astro-ph.CO]}}.

\bibitem{Sabti:2024jff}
N.~Sabti, R.~Reddy, J.~B. Mu\~noz, S.~Mishra-Sharma, and T.~Youn, ``{A Generative Modeling Approach to Reconstructing 21-cm Tomographic Data},'' \href{http://arxiv.org/abs/2407.21097}{{\ttfamily arXiv:2407.21097 [astro-ph.CO]}}.

\bibitem{Planck:2018vyg}
{\bfseries Planck} Collaboration, N.~Aghanim {\em et~al.}, ``{Planck 2018 results. VI. Cosmological parameters},'' \href{http://dx.doi.org/10.1051/0004-6361/201833910}{{\em Astron. Astrophys.} {\bfseries 641} (2020) A6}, \href{http://arxiv.org/abs/1807.06209}{{\ttfamily arXiv:1807.06209 [astro-ph.CO]}}. [Erratum: Astron.Astrophys. 652, C4 (2021)].

\bibitem{Mesinger2011:21cmFAST}
A.~{Mesinger}, S.~{Furlanetto}, and R.~{Cen}, ``{21CMFAST: a fast, seminumerical simulation of the high-redshift 21-cm signal},'' \href{http://dx.doi.org/10.1111/j.1365-2966.2010.17731.x}{{\em \mnras} {\bfseries 411} no.~2, (Feb., 2011) 955--972}, \href{http://arxiv.org/abs/1003.3878}{{\ttfamily arXiv:1003.3878 [astro-ph.CO]}}.

\bibitem{Furlanetto:2004sim}
S.~R. {Furlanetto}, M.~{Zaldarriaga}, and L.~{Hernquist}, ``{The Growth of H II Regions During Reionization},'' \href{http://dx.doi.org/10.1086/423025}{{\em ApJ} {\bfseries 613} no.~1, (Sept., 2004) 1--15}, \href{http://arxiv.org/abs/astro-ph/0403697}{{\ttfamily arXiv:astro-ph/0403697 [astro-ph]}}.

\bibitem{Park:2018ljd}
J.~Park, A.~Mesinger, B.~Greig, and N.~Gillet, ``{Inferring the astrophysics of reionization and cosmic dawn from galaxy luminosity functions and the 21-cm signal},'' \href{http://dx.doi.org/10.1093/mnras/stz032}{{\em Mon. Not. Roy. Astron. Soc.} {\bfseries 484} no.~1, (2019) 933--949}, \href{http://arxiv.org/abs/1809.08995}{{\ttfamily arXiv:1809.08995 [astro-ph.GA]}}.

\bibitem{jax2018github}
J.~Bradbury, R.~Frostig, P.~Hawkins, M.~J. Johnson, C.~Leary, D.~Maclaurin, G.~Necula, A.~Paszke, J.~Vander{P}las, S.~Wanderman-{M}ilne, and Q.~Zhang, ``{JAX}: composable transformations of {P}ython+{N}um{P}y programs,'' 2018.
\newblock \url{http://github.com/jax-ml/jax}.

\bibitem{Robnik:2022bzs}
J.~Robnik, G.~B. De~Luca, E.~Silverstein, and U.~Seljak, ``{Microcanonical Hamiltonian Monte Carlo},'' \href{http://arxiv.org/abs/2212.08549}{{\ttfamily arXiv:2212.08549 [stat.CO]}}.

\bibitem{Robnik:2023pgt}
J.~Robnik and U.~Seljak, ``{Fluctuation without dissipation: Microcanonical Langevin Monte Carlo},'' \href{http://arxiv.org/abs/2303.18221}{{\ttfamily arXiv:2303.18221 [hep-lat]}}.

\bibitem{nocedal1999numerical}
J.~Nocedal and S.~J. Wright, {\em Numerical optimization}.
\newblock Springer, 1999.

\bibitem{Schmittfull:2014tca}
M.~Schmittfull, T.~Baldauf, and U.~Seljak, ``{Near optimal bispectrum estimators for large-scale structure},'' \href{http://dx.doi.org/10.1103/PhysRevD.91.043530}{{\em Phys. Rev. D} {\bfseries 91} no.~4, (2015) 043530}, \href{http://arxiv.org/abs/1411.6595}{{\ttfamily arXiv:1411.6595 [astro-ph.CO]}}.

\bibitem{MoradinezhadDizgah:2019xun}
A.~Moradinezhad~Dizgah, H.~Lee, M.~Schmittfull, and C.~Dvorkin, ``{Capturing non-Gaussianity of the large-scale structure with weighted skew-spectra},'' \href{http://dx.doi.org/10.1088/1475-7516/2020/04/011}{{\em JCAP} {\bfseries 04} (2020) 011}, \href{http://arxiv.org/abs/1911.05763}{{\ttfamily arXiv:1911.05763 [astro-ph.CO]}}.

\bibitem{Chen:2024bdg}
S.-F. Chen, P.~Chakraborty, and C.~Dvorkin, ``{Analysis of BOSS galaxy data with weighted skew-spectra},'' \href{http://dx.doi.org/10.1088/1475-7516/2024/05/011}{{\em JCAP} {\bfseries 05} (2024) 011}, \href{http://arxiv.org/abs/2401.13036}{{\ttfamily arXiv:2401.13036 [astro-ph.CO]}}.

\bibitem{2015arXiv150504597R}
O.~{Ronneberger}, P.~{Fischer}, and T.~{Brox}, ``{U-Net: Convolutional Networks for Biomedical Image Segmentation},'' \href{http://dx.doi.org/10.48550/arXiv.1505.04597}{{\em arXiv e-prints} (May, 2015) arXiv:1505.04597}, \href{http://arxiv.org/abs/1505.04597}{{\ttfamily arXiv:1505.04597 [cs.CV]}}.

\bibitem{2016cvpr.confE...1H}
K.~{He}, X.~{Zhang}, S.~{Ren}, and J.~{Sun}, \href{http://dx.doi.org/10.1109/CVPR.2016.90}{``{Deep Residual Learning for Image Recognition},''} in {\em 2016 IEEE Conference on Computer Vision and Pattern Recognition (CVPR}, p.~1.
\newblock June, 2016.
\newblock \href{http://arxiv.org/abs/1512.03385}{{\ttfamily arXiv:1512.03385 [cs.CV]}}.

\bibitem{2018arXiv180308494W}
Y.~{Wu} and K.~{He}, ``{Group Normalization},'' \href{http://dx.doi.org/10.48550/arXiv.1803.08494}{{\em arXiv e-prints} (Mar., 2018) arXiv:1803.08494}, \href{http://arxiv.org/abs/1803.08494}{{\ttfamily arXiv:1803.08494 [cs.CV]}}.

\bibitem{2017arXiv171105101L}
I.~{Loshchilov} and F.~{Hutter}, ``{Decoupled Weight Decay Regularization},'' \href{http://dx.doi.org/10.48550/arXiv.1711.05101}{{\em arXiv e-prints} (Nov., 2017) arXiv:1711.05101}, \href{http://arxiv.org/abs/1711.05101}{{\ttfamily arXiv:1711.05101 [cs.LG]}}.

\bibitem{2016arXiv160803983L}
I.~{Loshchilov} and F.~{Hutter}, ``{SGDR: Stochastic Gradient Descent with Warm Restarts},'' \href{http://dx.doi.org/10.48550/arXiv.1608.03983}{{\em arXiv e-prints} (Aug., 2016) arXiv:1608.03983}, \href{http://arxiv.org/abs/1608.03983}{{\ttfamily arXiv:1608.03983 [cs.LG]}}.

\bibitem{MacKay2025:RULES}
V.~MacKay, Z.~Xu, R.~Byrne, and J.~Hewitt, ``{Complete Sampling of the $uv$ Plane with Realistic Radio Arrays: Introducing the RULES Algorithm, with Application to 21 cm Foreground Wedge Removal},'' \href{http://arxiv.org/abs/2509.15296}{{\ttfamily arXiv:2509.15296 [astro-ph.CO]}}.

\bibitem{Mertens2025:LOFAR_limits}
F.~G. {Mertens}, M.~{Mevius}, L.~V.~E. {Koopmans}, A.~R. {Offringa}, S.~{Zaroubi}, A.~{Acharya}, S.~A. {Brackenhoff}, E.~{Ceccotti}, E.~{Chapman}, K.~{Chege}, B.~{Ciardi}, R.~{Ghara}, S.~{Ghosh}, S.~K. {Giri}, I.~{Hothi}, C.~{H{\"o}fer}, I.~T. {Iliev}, V.~{Jeli{\'c}}, Q.~{Ma}, G.~{Mellema}, S.~{Munshi}, V.~N. {Pandey}, and S.~{Yatawatta}, ``{Deeper multi-redshift upper limits on the epoch of reionisation 21 cm signal power spectrum from LOFAR between z = 8.3 and z = 10.1},'' \href{http://dx.doi.org/10.1051/0004-6361/202554158}{{\em \aap} {\bfseries 698} (June, 2025) A186}, \href{http://arxiv.org/abs/2503.05576}{{\ttfamily arXiv:2503.05576 [astro-ph.CO]}}.

\bibitem{Ceccotti2025:LOFAR_limit}
E.~{Ceccotti}, A.~R. {Offringa}, F.~G. {Mertens}, L.~V.~E. {Koopmans}, S.~{Munshi}, J.~K. {Chege}, A.~{Acharya}, S.~A. {Brackenhoff}, E.~{Chapman}, B.~{Ciardi}, R.~{Ghara}, S.~{Ghosh}, S.~K. {Giri}, C.~{H{\"o}fer}, I.~{Hothi}, G.~{Mellema}, M.~{Mevius}, V.~N. {Pandey}, and S.~{Zaroubi}, ``{First upper limits on the 21-cm signal power spectrum of neutral hydrogen at $z=9.16$ from the LOFAR 3C196 field},'' \href{http://dx.doi.org/10.48550/arXiv.2504.18534}{{\em arXiv e-prints} (Apr., 2025) arXiv:2504.18534}, \href{http://arxiv.org/abs/2504.18534}{{\ttfamily arXiv:2504.18534 [astro-ph.CO]}}.

\bibitem{Trott2020:MWA_Limit}
C.~M. {Trott}, C.~H. {Jordan}, S.~{Midgley}, N.~{Barry}, B.~{Greig}, B.~{Pindor}, J.~H. {Cook}, G.~{Sleap}, S.~J. {Tingay}, D.~{Ung}, P.~{Hancock}, A.~{Williams}, J.~{Bowman}, R.~{Byrne}, A.~{Chokshi}, B.~J. {Hazelton}, K.~{Hasegawa}, D.~{Jacobs}, R.~C. {Joseph}, W.~{Li}, J.~{Line}, C.~{Lynch}, B.~{McKinley}, D.~A. {Mitchell}, M.~F. {Morales}, M.~{Ouchi}, J.~C. {Pober}, M.~{Rahimi}, K.~{Takahashi}, R.~B. {Wayth}, R.~L. {Webster}, M.~{Wilensky}, J.~S.~B. {Wyithe}, S.~{Yoshiura}, Z.~{Zhang}, and Q.~{Zheng}, ``{Deep multi-redshift limits on Epoch of Reionisation 21 cm Power Spectra from Four Seasons of Murchison Widefield Array Observations},'' \href{http://dx.doi.org/10.1093/mnras/staa414}{{\em \mnras} (Feb., 2020) }, \href{http://arxiv.org/abs/2002.02575}{{\ttfamily arXiv:2002.02575 [astro-ph.CO]}}.

\bibitem{Nunhokee2025:MWA_limit}
C.~D. {Nunhokee}, D.~{Null}, C.~M. {Trott}, N.~{Barry}, Y.~{Qin}, R.~B. {Wayth}, J.~L.~B. {Line}, C.~H. {Jordan}, B.~{Pindor}, J.~H. {Cook}, J.~{Bowman}, A.~{Chokshi}, J.~{Ducharme}, K.~{Elder}, Q.~{Guo}, B.~{Hazelton}, W.~{Hidayat}, T.~{Ito}, D.~{Jacobs}, E.~{Jong}, M.~{Kolopanis}, T.~{Kunicki}, E.~{Lilleskov}, M.~F. {Morales}, J.~C. {Pober}, A.~{Selvaraj}, R.~{Shi}, K.~{Takahashi}, S.~J. {Tingay}, R.~L. {Webster}, S.~{Yoshiura}, and Q.~{Zheng}, ``{Limits on the 21 cm Power Spectrum at z = 6.5{\textendash}7.0 from Murchison Widefield Array Observations},'' \href{http://dx.doi.org/10.3847/1538-4357/adda45}{{\em \apj} {\bfseries 989} no.~1, (Aug., 2025) 57}, \href{http://arxiv.org/abs/2505.09097}{{\ttfamily arXiv:2505.09097 [astro-ph.CO]}}.

\bibitem{HERA2022:h1c_idr2_limit}
Z.~{Abdurashidova}, J.~E. {Aguirre}, P.~{Alexander}, Z.~S. {Ali}, Y.~{Balfour}, A.~P. {Beardsley}, G.~{Bernardi}, T.~S. {Billings}, J.~D. {Bowman}, R.~F. {Bradley}, P.~{Bull}, J.~{Burba}, S.~{Carey}, C.~L. {Carilli}, C.~{Cheng}, D.~R. {DeBoer}, M.~{Dexter}, E.~{de Lera Acedo}, T.~{Dibblee-Barkman}, J.~S. {Dillon}, J.~{Ely}, A.~{Ewall-Wice}, N.~{Fagnoni}, R.~{Fritz}, S.~R. {Furlanetto}, K.~{Gale-Sides}, B.~{Glendenning}, D.~{Gorthi}, B.~{Greig}, J.~{Grobbelaar}, Z.~{Halday}, B.~J. {Hazelton}, J.~N. {Hewitt}, J.~{Hickish}, D.~C. {Jacobs}, A.~{Julius}, N.~S. {Kern}, J.~{Kerrigan}, P.~{Kittiwisit}, S.~A. {Kohn}, M.~{Kolopanis}, A.~{Lanman}, P.~{La Plante}, T.~{Lekalake}, D.~{Lewis}, A.~{Liu}, D.~{MacMahon}, L.~{Malan}, C.~{Malgas}, M.~{Maree}, Z.~E. {Martinot}, E.~{Matsetela}, A.~{Mesinger}, M.~{Molewa}, M.~F. {Morales}, T.~{Mosiane}, S.~G. {Murray}, A.~R. {Neben}, B.~{Nikolic}, C.~D. {Nunhokee}, A.~R. {Parsons}, N.~{Patra}, R.~{Pascua}, S.~{Pieterse}, J.~C. {Pober}, N.~{Razavi-Ghods}, J.~{Ringuette},
  J.~{Robnett}, K.~{Rosie}, P.~{Sims}, S.~{Singh}, C.~{Smith}, A.~{Syce}, N.~{Thyagarajan}, P.~K.~G. {Williams}, H.~{Zheng}, and {HERA Collaboration}, ``{First Results from HERA Phase I: Upper Limits on the Epoch of Reionization 21 cm Power Spectrum},'' \href{http://dx.doi.org/10.3847/1538-4357/ac1c78}{{\em \apj} {\bfseries 925} no.~2, (Feb., 2022) 221}, \href{http://arxiv.org/abs/2108.02263}{{\ttfamily arXiv:2108.02263 [astro-ph.CO]}}.

\bibitem{HERA2023:h1c_idr3_limit}
{HERA Collaboration}, Z.~{Abdurashidova}, T.~{Adams}, J.~E. {Aguirre}, P.~{Alexander}, Z.~S. {Ali}, R.~{Baartman}, Y.~{Balfour}, R.~{Barkana}, A.~P. {Beardsley}, G.~{Bernardi}, T.~S. {Billings}, J.~D. {Bowman}, R.~F. {Bradley}, D.~{Breitman}, P.~{Bull}, J.~{Burba}, S.~{Carey}, C.~L. {Carilli}, C.~{Cheng}, S.~{Choudhuri}, D.~R. {DeBoer}, E.~{de Lera Acedo}, M.~{Dexter}, J.~S. {Dillon}, J.~{Ely}, A.~{Ewall-Wice}, N.~{Fagnoni}, A.~{Fialkov}, R.~{Fritz}, S.~R. {Furlanetto}, K.~{Gale-Sides}, H.~{Garsden}, B.~{Glendenning}, A.~{Gorce}, D.~{Gorthi}, B.~{Greig}, J.~{Grobbelaar}, Z.~{Halday}, B.~J. {Hazelton}, S.~{Heimersheim}, J.~N. {Hewitt}, J.~{Hickish}, D.~C. {Jacobs}, A.~{Julius}, N.~S. {Kern}, J.~{Kerrigan}, P.~{Kittiwisit}, S.~A. {Kohn}, M.~{Kolopanis}, A.~{Lanman}, P.~{La Plante}, D.~{Lewis}, A.~{Liu}, A.~{Loots}, Y.-Z. {Ma}, D.~H.~E. {MacMahon}, L.~{Malan}, K.~{Malgas}, C.~{Malgas}, M.~{Maree}, B.~{Marero}, Z.~E. {Martinot}, L.~{McBride}, A.~{Mesinger}, J.~{Mirocha}, M.~{Molewa}, M.~F. {Morales},
  T.~{Mosiane}, J.~B. {Mu{\~n}oz}, S.~G. {Murray}, V.~{Nagpal}, A.~R. {Neben}, B.~{Nikolic}, C.~D. {Nunhokee}, H.~{Nuwegeld}, A.~R. {Parsons}, R.~{Pascua}, N.~{Patra}, S.~{Pieterse}, Y.~{Qin}, N.~{Razavi-Ghods}, J.~{Robnett}, K.~{Rosie}, M.~G. {Santos}, P.~{Sims}, S.~{Singh}, C.~{Smith}, H.~{Swarts}, J.~{Tan}, N.~{Thyagarajan}, M.~J. {Wilensky}, P.~K.~G. {Williams}, P.~{van Wyngaarden}, and H.~{Zheng}, ``{Improved Constraints on the 21 cm EoR Power Spectrum and the X-Ray Heating of the IGM with HERA Phase I Observations},'' \href{http://dx.doi.org/10.3847/1538-4357/acaf50}{{\em \apj} {\bfseries 945} no.~2, (Mar., 2023) 124}.

\bibitem{HERA2025:PhaseII_Limit}
{\bfseries HERA} Collaboration, {HERA Collaboration}, Z.~{Abdurashidova}, T.~{Adams}, J.~E. {Aguirre}, R.~{Baartman}, R.~{Barkana}, L.~M. {Berkhout}, G.~{Bernardi}, T.~S. {Billings}, B.~B. {Bizarria}, J.~D. {Bowman}, D.~{Breitman}, P.~{Bull}, J.~{Burba}, R.~{Byrne}, S.~{Carey}, R.~S. {Chandra}, K.-F. {Chen}, S.~{Choudhuri}, T.~{Cox}, D.~R. {DeBoer}, E.~{de Lera Acedo}, M.~{Dexter}, J.~{Dhandha}, J.~S. {Dillon}, S.~{Dynes}, N.~{Eksteen}, J.~{Ely}, A.~{Ewall-Wice}, N.~{Fagnoni}, A.~{Fialkov}, S.~R. {Furlanetto}, K.~{Gale-Sides}, H.~{Garsden}, A.~{Gorce}, D.~{Gorthi}, Z.~{Halday}, B.~J. {Hazelton}, J.~N. {Hewitt}, J.~{Hickish}, T.~{Huang}, D.~C. {Jacobs}, A.~{Josaitis}, N.~S. {Kern}, J.~{Kerrigan}, P.~{Kittiwisit}, M.~{Kolopanis}, A.~{Lanman}, P.~{La Plante}, A.~{Liu}, Y.-Z. {Ma}, D.~H.~E. {MacMahon}, L.~{Malan}, C.~{Malgas}, K.~{Malgas}, B.~{Marero}, Z.~E. {Martinot}, L.~{McBride}, A.~{Mesinger}, J.~{Mirocha}, N.~{Mohamed-Hinds}, M.~{Molewa}, M.~F. {Morales}, J.~B. {Mu{\~n}oz}, S.~G. {Murray}, B.~{Nikolic},
  H.~{Nuwegeld}, A.~R. {Parsons}, R.~{Pascua}, N.~{Patra}, S.~{Pochinda}, Y.~{Qin}, E.~{Rath}, N.~{Razavi-Ghods}, D.~{Riley}, K.~{Rosie}, M.~G. {Santos}, S.~{Singh}, D.~{Storer}, H.~{Swarts}, J.~{Tan}, E.~{Th{\'e}lie}, P.~{van Wyngaarden}, M.~J. {Wilensky}, P.~K.~G. {Williams}, and H.~{Zheng}, ``{First Results from HERA Phase II},'' \href{http://arxiv.org/abs/2511.21289}{{\ttfamily arXiv:2511.21289 [astro-ph.CO]}}.

\bibitem{Hutter2017:21cm_LAE}
A.~{Hutter}, P.~{Dayal}, V.~{M{\"u}ller}, and C.~M. {Trott}, ``{Exploring 21cm-Lyman Alpha Emitter Synergies for SKA},'' \href{http://dx.doi.org/10.3847/1538-4357/836/2/176}{{\em \apj} {\bfseries 836} no.~2, (Feb., 2017) 176}, \href{http://arxiv.org/abs/1605.01734}{{\ttfamily arXiv:1605.01734 [astro-ph.CO]}}.

\bibitem{Heneka2021:21cm_LAE}
C.~{Heneka} and A.~{Cooray}, ``{Optimal survey parameters: Ly {\ensuremath{\alpha}} and H {\ensuremath{\alpha}} intensity mapping for synergy with the 21-cm signal during reionization},'' \href{http://dx.doi.org/10.1093/mnras/stab1842}{{\em \mnras} {\bfseries 506} no.~2, (Sept., 2021) 1573--1584}, \href{http://arxiv.org/abs/2104.12739}{{\ttfamily arXiv:2104.12739 [astro-ph.CO]}}.

\bibitem{Davies2021:21cm_stack}
J.~E. {Davies}, R.~A.~C. {Croft}, T.~{Di-Matteo}, B.~{Greig}, Y.~{Feng}, and J.~S.~B. {Wyithe}, ``{Stacking redshifted 21 cm images of H II regions around high-redshift galaxies as a probe of early reionization},'' \href{http://dx.doi.org/10.1093/mnras/staa3531}{{\em \mnras} {\bfseries 501} no.~1, (Feb., 2021) 146--156}, \href{http://arxiv.org/abs/2007.13318}{{\ttfamily arXiv:2007.13318 [astro-ph.CO]}}.

\bibitem{Trott2021:MWA_LAEs}
C.~M. {Trott}, C.~H. {Jordan}, J.~L.~B. {Line}, C.~R. {Lynch}, S.~{Yoshiura}, B.~{McKinley}, P.~{Dayal}, B.~{Pindor}, A.~{Hutter}, K.~{Takahashi}, R.~B. {Wayth}, N.~{Barry}, A.~{Beardsley}, J.~{Bowman}, R.~{Byrne}, A.~{Chokshi}, B.~{Greig}, K.~{Hasegawa}, B.~J. {Hazelton}, E.~{Howard}, D.~{Jacobs}, M.~{Kolopanis}, D.~A. {Mitchell}, M.~F. {Morales}, S.~{Murray}, J.~C. {Pober}, M.~{Rahimi}, S.~J. {Tingay}, R.~L. {Webster}, M.~{Wilensky}, J.~S.~B. {Wyithe}, and Q.~{Zheng}, ``{Constraining the 21 cm brightness temperature of the IGM at z = 6.6 around LAEs with the murchison widefield array},'' \href{http://dx.doi.org/10.1093/mnras/stab2235}{{\em \mnras} {\bfseries 507} no.~1, (Oct., 2021) 772--780}, \href{http://arxiv.org/abs/2107.14493}{{\ttfamily arXiv:2107.14493 [astro-ph.CO]}}.

\bibitem{Cox2022:cross-correlation}
T.~A. {Cox}, D.~C. {Jacobs}, and S.~G. {Murray}, ``{Estimating the feasibility of 21cm-Ly{\ensuremath{\alpha}} synergies using the hydrogen Epoch of Reionization array},'' \href{http://dx.doi.org/10.1093/mnras/stac486}{{\em \mnras} {\bfseries 512} no.~1, (May, 2022) 792--801}, \href{http://arxiv.org/abs/2202.08957}{{\ttfamily arXiv:2202.08957 [astro-ph.CO]}}.

\bibitem{LaPlante2023:21cm_x_Roman}
P.~{La Plante}, J.~{Mirocha}, A.~{Gorce}, A.~{Lidz}, and A.~{Parsons}, ``{Prospects for 21 cm Galaxy Cross-correlations with HERA and the Roman High-latitude Survey},'' \href{http://dx.doi.org/10.3847/1538-4357/acaeb0}{{\em \apj} {\bfseries 944} no.~1, (Feb., 2023) 59}, \href{http://arxiv.org/abs/2205.09770}{{\ttfamily arXiv:2205.09770 [astro-ph.CO]}}.

\bibitem{Gagnon-Hartman2025:21cm_x_LAE}
S.~{Gagnon-Hartman}, J.~{Davies}, and A.~{Mesinger}, ``{Detecting galaxy-21-cm cross-correlation during reionization},'' \href{http://dx.doi.org/10.48550/arXiv.2502.20447}{{\em arXiv e-prints} (Feb., 2025) arXiv:2502.20447}, \href{http://arxiv.org/abs/2502.20447}{{\ttfamily arXiv:2502.20447 [astro-ph.CO]}}.

\bibitem{Hutter2025:Cross_Correlation}
A.~{Hutter} and C.~{Heneka}, ``{The 21cm-galaxy cross-correlation: Realistic forecast for 21cm signal detection and reionisation constraints},'' \href{http://dx.doi.org/10.48550/arXiv.2509.15906}{{\em arXiv e-prints} (Sept., 2025) arXiv:2509.15906}, \href{http://arxiv.org/abs/2509.15906}{{\ttfamily arXiv:2509.15906 [astro-ph.CO]}}.

\bibitem{Chen2025:21cm_x_LAEs}
K.-F. {Chen}, M.~{Neyer}, J.~N. {Hewitt}, A.~{Smith}, and M.~{Vogelsberger}, ``{Stacking 21-cm Maps around Lyman-$\alpha$ Emitters during Reionization: Prospects for a Cross-correlation Detection with the Hydrogen Epoch of Reionization Array},'' \href{http://dx.doi.org/10.48550/arXiv.2510.07374}{{\em arXiv e-prints} (Oct., 2025) arXiv:2510.07374}, \href{http://arxiv.org/abs/2510.07374}{{\ttfamily arXiv:2510.07374 [astro-ph.CO]}}.

\bibitem{Kern:2025bqm}
N.~Kern, ``{A Differentiable, End-to-End Forward Model for 21 cm Cosmology: Estimating the Foreground, Instrument, and Signal Joint Posterior},'' \href{http://arxiv.org/abs/2504.07090}{{\ttfamily arXiv:2504.07090 [astro-ph.CO]}}.

\bibitem{Michaux:2020yis}
M.~Michaux, O.~Hahn, C.~Rampf, and R.~E. Angulo, ``{Accurate initial conditions for cosmological N-body simulations: Minimizing truncation and discreteness errors},'' \href{http://dx.doi.org/10.1093/mnras/staa3149}{{\em Mon. Not. Roy. Astron. Soc.} {\bfseries 500} no.~1, (2020) 663--683}, \href{http://arxiv.org/abs/2008.09588}{{\ttfamily arXiv:2008.09588 [astro-ph.CO]}}.

\bibitem{Jing:2004fq}
Y.~P. Jing, ``{Correcting for the alias effect when measuring the power spectrum using FFT},'' \href{http://dx.doi.org/10.1086/427087}{{\em Astrophys. J.} {\bfseries 620} (2005) 559--563}, \href{http://arxiv.org/abs/astro-ph/0409240}{{\ttfamily arXiv:astro-ph/0409240}}.

\end{thebibliography}\endgroup
\end{document}